\documentclass[pra,twocolumn,aps,superscriptaddress]{revtex4}
\usepackage{graphics}
\usepackage{bm}
\usepackage[fleqn]{amsmath}
\usepackage{hyperref}
\usepackage{exscale,relsize}
\hypersetup{
     colorlinks=true,       
    linkcolor=blue,          
    citecolor=blue,        
    filecolor=magenta,      
    urlcolor=cyan           
}
\usepackage{epsfig}
\usepackage{amssymb}
\usepackage{amsmath}
\usepackage{euscript}
\usepackage{float}
 \usepackage{amsthm}

\newcommand{\be}{\begin{equation}}
\newcommand{\e}{\end{equation}}

\newcommand{\beml}{\begin{subequations}}
\newcommand{\eml}{\end{subequations}}
\newcommand{\beq}{\begin{eqnarray}}
\newcommand{\eq}{\end{eqnarray}}
\newcommand{\ba}{\begin{array}}
\newcommand{\ea}{\end{array}}

\newcommand{\lt}{\left}
\newcommand{\rt}{\right}
\newcommand{\n}{\nonumber}

\newcommand{\s}{\sigma}
\newcommand{\la}{\langle}
\newcommand{\ra}{\rangle}

\newcommand{\avg}[1]{\left< #1 \right>} 
\newcommand{\avrg}[1]{\langle #1 \rangle}

\newcommand{\ket}[1]{\left| #1 \right>} 
\newcommand{\bra}[1]{\left< #1 \right|} 
\let\baraccent=\= 
\renewcommand{\=}[1]{\stackrel{#1}{=}} 

\theoremstyle{definition}

\theoremstyle{remark}

\begin{document}

\title{Coherent backscattering of intense light by cold atoms with degenerate energy levels: Diagrammatic treatment}

\author{V.N. Shatokhin}
\affiliation{Institute of Physics, University of Freiburg, Hermann-Herder-Strasse 3, D-79104 Freiburg, Germany}
\author{R. Blattmann}
\affiliation{Institute of Physics, University of Freiburg, Hermann-Herder-Strasse 3, D-79104 Freiburg, Germany}
\affiliation{Institute of Physics, University of Augsburg, Universit\"atsstr. 1,
D-86135 Augsburg, Germany}
\author{T. Wellens}
\affiliation{Institute of Physics, University of Freiburg, Hermann-Herder-Strasse 3, D-79104 Freiburg, Germany}
\author{A. Buchleitner}
\affiliation{Institute of Physics, University of Freiburg, Hermann-Herder-Strasse 3, D-79104 Freiburg, Germany}

\date{\today}

\begin{abstract}
We present a generalization of the diagrammatic pump-probe approach
to coherent backscattering (CBS) of intense laser light for atoms
with degenerate energy levels. We employ this approach for a
characterization of the double scattering signal from optically
pumped atoms with the transition $J_g\rightarrow J_e=J_g+1$ in the
helicity preserving polarization channel. We show that, in the
saturation regime, the internal degeneracy becomes manifest for
atoms with $J_g\geq 1$, leading to a faster decrease of the CBS
enhancement factor with increasing saturation parameter than in the
non-degenerate case.
\end{abstract}


\maketitle



\section{Introduction}

Coherent backscattering (CBS) is an interference phenomenon arising when monochromatic waves get multiply
scattered by a disordered
distribution of dilute scatterers. It occurs in the weak scattering regime, where the
constructive interference of counter-propagating amplitudes survives the disorder average and leads to
an enhanced intensity in backscattering direction \cite{Lagendijk96,sheng}.
CBS was observed for the first time with optical waves and polystyrene particles
acting as classical point scatters \cite{albada85} in the 80's, and, more recently, with acoustic \cite{bayer93}, seismic \cite{larose04}, and matter \cite{josse12} waves.

Constant technical progress and modern cooling techniques made it
possible to study CBS of laser light by clouds of cold atoms
behaving like unique quantum scatterers \cite{labeyrie99}. In contrast
to classical scatterers, atoms are able to scatter light
inelastically, when driven by an  intense resonant laser field
\cite{cohen-tannoudji}. Moreover, the electronic structure of the
atoms allows the scattered photons to flip their polarization and renders the scattering process polarization-dependent. Recent
experiments on CBS of light by cold Sr \cite{chaneliere04} and Rb
\cite{balik05} atoms showed that nonlinear inelastic scattering, as
well as the internal atomic structure strongly affect the phase
coherence of the multiply scattered fields, reducing the
interference contrast. However, an accurate quantitative description
of the above experiments is still missing. Indeed, the theoretical
approaches using a diagrammatic scattering theory \cite{wellens04},
a master equation \cite{shatokhin05}, or quantum Langevin equations
\cite{gremaud06} led to a deeper understanding of the physical
mechanism responsible for the observed coherence loss and achieved a
qualitative agreement with the experiment \cite{chaneliere04}, but
were unable to reach an accurate quantitative description thereof.
The major problem with the above approaches is that they are
restricted either to a small number of photons or atoms. In
particular, the master equation approach is capable of accurately
assessing the atomic response to a strong resonant field, but the
complexity of the problem increases exponentially with the number of
atoms.

Recently, we suggested a hybrid -- diagrammatic pump-probe --
approach, which blends diagrammatic scattering theory and single-atom master equations (or optical
Bloch equations (OBE))
\cite{geiger10,wellens10}. This method was initially introduced for the double
scattering contribution to CBS from two two-level atoms, in which
case the signal is deduced from solutions of the OBE under a
classical bichromatic driving. One component of the bichromatic
driving represents the in general saturating laser field, while the other, non-saturating component stems from the field scattered by the
second atom. Thus, our approach owes its name to the analogy with a
method in saturation spectroscopy \cite{meystre07}.

First of all, since the diagrammatic pump-probe approach uses only
single-atom quantities for the derivation of the multiple scattering
signal, it circumvents the aforementioned problem of the exponential
growth of the system complexity with the number of scatterers.
Second, it transforms the problem of CBS of intense laser light off
a cold atomic cloud into a form that is amenable to Monte-Carlo
simulation methods \cite{binninger12}. Third, for double
\cite{shatokhin10} and triple \cite{shatokhin12b} scattering orders
the solutions obtained within the pump-probe approach are equivalent
to the solutions following from the master equation (where, in the triple scattering case, the recurrent scattering contributions are dropped). Under the
assumption that this equivalence always holds in the dilute regime,
general analytical expressions have recently been derived for single-atom responses \cite{shatokhin12b}. It will be a subject of future
work to include these expressions into the Monte-Carlo simulation
subroutines.

The purpose of the present contribution is to generalize the diagrammatic pump-probe
approach to realistic dipole transitions possessing internal degeneracy. Such
transitions were probed in the above-mentioned experiments \cite{chaneliere04,balik05}. We will also
incorporate a vectorial representation of the electromagnetic fields into our approach,
which is required for a proper description of the light-matter interaction as well as of the
polarization-sensitive character of the CBS effect.

The paper is structured as follows. In the next section, we recall
the basic ingredients of the pump-probe approach for two-level
atoms. In Sec.~\ref{sec:vector}, we generalize this approach to the
scenario of vector fields and atoms with degenerate dipole
transitions. Thereafter, we apply this generalized treatment to
double scattering from optically pumped atoms with the ground and
excited state angular momenta $J_g$ and $J_e=J_g+1$, respectively,
in the helicity preserving polarization channel. We show that the
{\it elastic} component of the double scattering spectrum for
arbitrary $J_g$ can be expressed using the results for $J_g=0$. This
is {\it not} in general the case for the {\it inelastic} intensity,
since inelastic scattering from the degenerate ground state
results in additional processes that do not interfere perfectly, and
lead to a more rapid decay of phase coherence as compared to atoms
with $J_g=0$. Finally, in Sec.~\ref{sec:conclusion} we conclude our
work.

\section{The diagrammatic pump-probe approach for two-level atoms}
\label{sec:two-level}

Before we present the pump-probe approach to CBS from two atoms with
degenerate energy levels, it is instructive to recall its
formulation for two-level atoms \cite{geiger10, wellens10}. The
generalization thereof for multilevel dipole transitions will be
developed, along the same lines, in Sec.~\ref{sec:vector}.

\begin{figure}
\includegraphics[width=8cm]{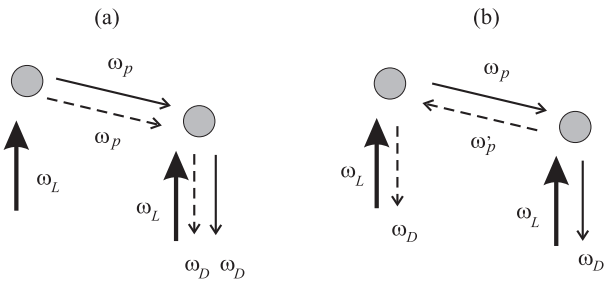}
\caption{Double scattering processes surviving the disorder average:
(a) ladder, or background, contribution, describing co-propagating
amplitudes; (b) crossed, or interference, contribution, resulting
from the interference between counter-propagating amplitudes.
The incident laser field at frequency $\omega_L$ is assumed to be
strong enough to induce nonlinear inelastic scattering processes,
whereupon the frequencies $\omega_p$, $\omega_p^\prime$, and
$\omega_D$ may differ from the incident laser field's frequency. }
\label{fig:processes}
\end{figure}

To this end, let us consider double scattering in a toy model of
CBS, consisting of two immobile and distant atoms in free space,
driven by a near-resonant laser field. The scattering processes
which survive the disorder average and contribute to the background
and interference intensities, respectively, are shown in
Fig.~\ref{fig:processes}(a) and (b). Thick arrows directed towards
grey dots depict a cw (continuous wave) laser field of arbitrary
strength driving the atoms. Thin solid (dashed) arrows depict
positive- (negative-)frequency parts of the scattered field.
Now, the main idea of the pump-probe approach is to account for the laser-atom interaction non-perturbatively, while the
atom-atom interaction is dealt with perturbatively, at lowest
non-vanishing order \cite{gush74,agarwal86,mollow72}. The two
components of the driving field seen by each of the atoms in Fig. \ref{fig:processes} correspond to the incident laser
field and the field scattered by the other atom, respectively. A large interatomic
separation implies a small Rabi frequency of the scattered field in comparison to the natural line width, and justifies its perturbative treatment. As regards the classical ansatz for the probe field, it was
suggested \cite{geiger10, wellens10} and proven
\cite{shatokhin10} that it is valid up to second order in the
scattered field (two exchanged amplitudes), because the non-classical character of the atomic
radiation reveals itself only in the field correlation
functions describing the coincidence measurements of at least two
photons \cite{kimble77} (i.e., four exchanged amplitudes).

The classical description of the scattered fields allows us to
consider the light-matter interaction of each of the atoms
separately, and to derive the double-scattering signal by combining
single-atom building blocks, in analogy with multiple scattering
theory \cite{Lagendijk96}.

According to \cite{geiger10, wellens10}, the single-atom building
blocks describe stationary spectral responses of a two-level atom
subjected to a classical bichromatic electric field $E_{pp}(t)$:
 \begin{equation}
  \label{eq:1}
  E_{pp}(t)=\mathcal{E} e^{-i\omega_L
    t}+\mathcal{E}^*e^{i \omega_L t}+\epsilon e^{-i\omega_p t}+\epsilon^*
 e^{i\omega_p t},
\end{equation}
where both waves, whose frequencies are introduced in Fig.~\ref{fig:processes}, are split into their positive- and
negative-frequency parts, with $\mathcal{E}$ and $\epsilon$ being,
respectively, the complex amplitudes of the laser and the scattered
fields, the latter acting as a ``probe'' on the laser-driven atom.
Since CBS is observed in the dilute regime, i.e., when $k_Lr_{12}\gg
1$, the atoms are located in the radiation zone of each other where
the probe field scales as $(k_Lr_{12})^{-1}$, validating a
perturbative treatment.

The dynamics of the quantum-mechanical expectation value of an
arbitrary atomic observable $Q$ of a two-level atom in free space
driven by the classical field (\ref{eq:1}) can be deduced from a
standard master equation for single atom resonance fluorescence
under classical bichromatic driving, which in the frame rotating at the laser
frequency reads (see, for instance, \cite{agarwal86})
\begin{align}
\la\dot{Q}\ra&=\left\la-i\delta[\s^+\s^-,Q]
-\frac{i}{2}[\Omega \s^++\Omega^*\s^-,Q]\right.\n\\
&\left.+\gamma\lt(\s^+[Q,\s^-]+[\s^+,Q]\s^-\rt)\right.\n\\
&\left. -\frac{i}{2}[g e^{-i\omega t}\s^++g^* e^{i\omega
t}\s^-,Q]\right\ra. \label{eq:master}
\end{align}
Here, $\sigma^- (\sigma^+)
 = \ket{g}\bra{e} (\ket{e}\bra{g})$
denotes the atomic lowering (raising) operator, with $\ket{g}$ and
$\ket{e}$ the atomic ground and excited states, respectively.
Furthermore, $\delta=\omega_L-\omega_0$ is the detuning between the laser and the
atomic transition frequency, $\omega=\omega_p-\omega_L$ the detuning
between the probe and the laser field frequency, $\gamma$ half the
spontaneous decay rate of the excited state, and
$\Omega=2d\mathcal{E}/\hbar$, $g=2d\epsilon/\hbar$, with $d$ the
matrix element of the dipole transition, the Rabi frequencies of the
laser and probe fields, respectively.

Equation (\ref{eq:master}) is equivalent to the OBE with bichromatic driving, which we write
in matrix form as \cite{geiger10,wellens10}:
\begin{align}
   \label{eq:MOBE}
   \avrg{\dot{\boldsymbol{\sigma}}(t)}&={\bf M}_1
  \avg{\boldsymbol{\sigma}(t)}+{\bf L}_1+ge^{-i\omega
    t}\boldsymbol{\Delta}^{(-)}\avg{\boldsymbol{\sigma}(t)}\n\\
   &+g^*e^{i\omega
    t}\boldsymbol{\Delta}^{(+)}\avg{\boldsymbol{\sigma}(t)}.
 \end{align}
Here, $\avg{\boldsymbol{\sigma}}=(\avg{\sigma^-},\avg{\sigma^+},\avg{\sigma^z})$
is the quantum-mechanical expectation value of the optical Bloch
vector, with $\sigma^z=\sigma^+\sigma^- - \sigma^-\sigma^+$, and the explicit form of the matrices $\bf{M}_1$,
$\boldsymbol{\Delta}^{(+)}$, $\boldsymbol{\Delta}^{(-)}$, together with the vector $\bf{L}_1$, is readily obtained when the elements of the Bloch vector are entered into Eq.~(\ref{eq:master}).

The basic quantity that we are using to characterize single-atom
stationary spectral responses is the frequency correlation function,
\begin{align}
I(\nu,\nu^\prime)&=\frac{1}{(2\pi)^2}\int_{-\infty}^{\infty}dt\int_{-\infty}^{\infty}dt^\prime
e^{-i t \nu+i t^\prime \nu^\prime}\n\\
&\times \la\s^+(t)\s^-(t^\prime)\ra,
\label{eq:spec_corr_func}
\end{align}
which describes spectral correlations between the positive-frequency
amplitude at frequency $\nu$ and the negative-frequency
amplitude at frequency $\nu^\prime$. To evaluate this function, we
split the atomic dipole temporal correlation function
$\la\s^+(t)\s^-(t^\prime)\ra$ in Eq.~(\ref{eq:spec_corr_func}) into
a sum of a factorized and a fluctuating part, respectively:
\be
\la\s^+(t)\s^-(t^\prime)\ra=\la\s^+(t)\ra\la\s^-(t^\prime)\ra+
\la\Delta\s^+(t)\Delta\s^-(t^\prime)\ra, \label{eq:fact_and_fluct}
\e where $\Delta\s^\pm\equiv \s^\pm-\la\s^\pm\ra$. Insertion of the
right-hand side of Eq.~(\ref{eq:fact_and_fluct}) into
Eq.~(\ref{eq:spec_corr_func}) yields a decomposition \be
I(\nu,\nu^\prime)=I^{\rm el}(\nu,\nu^\prime)+I^{\rm
in}(\nu,\nu^\prime), \label{eq:int_el_in}\e where elastic and
inelastic components, $I^{\rm el}(\nu,\nu^\prime)$ and $I^{\rm
in}(\nu,\nu^\prime)$, result from the Fourier transform of the
factorized and the fluctuating part of the atomic dipole
correlation function (\ref{eq:fact_and_fluct}), respectively.

The stationary factorized atomic dipole correlation function, defined in terms of the atomic
raising and lowering operators, can readily be evaluated from the perturbative solutions of
Eq.~(\ref{eq:MOBE}) to second order in the probe field amplitude. A similar
consideration applies also to  the fluctuating part of the atomic dipole correlation function,
since, according to the quantum regression theorem \cite{cohen-tannoudji}, it satisfies an equation of motion
which follows straightforwardly from (\ref{eq:MOBE}).  Plugging the obtained solutions into
Eq.~(\ref{eq:spec_corr_func}), and performing the Fourier transformations, we obtain the
elastic and inelastic single atom-spectral responses.

In frequency space, the perturbative solutions for the atomic dipole averages and correlation functions
are referred to as the {\it elementary} single-atoms building blocks.
\begin{figure}
\includegraphics[width=8cm]{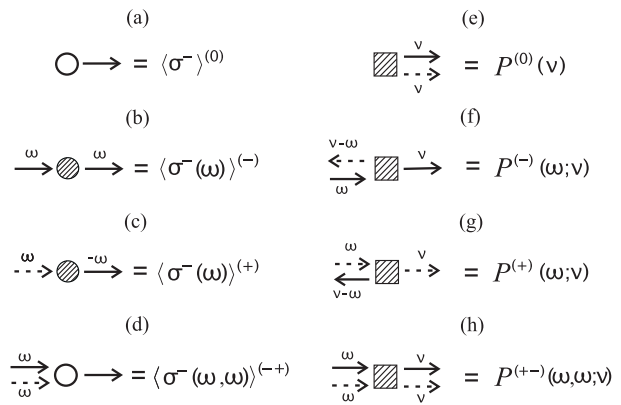}
\caption{Graphical definitions of the elementary single-atom spectral responses, together with our
notation for the corresponding correlation functions. (a)-(d)  complex scattering amplitudes
associated with the perturbative solutions of Eq.~(\ref{eq:MOBE}); (e)-(h)
blocks associated with the perturbative solutions for the fluctuating part of the atomic
dipole correlation function
$\la\Delta\s^+(t)\Delta\s^-(t^\prime)\ra$ (see Eq.~(\ref{eq:fact_and_fluct})). Blank and hatched shapes denote elastic and inelastic spectral responses, respectively (see text).} \label{fig:double}
\end{figure}
It is convenient to define them graphically \cite{shatokhin12a}.
Figure \ref{fig:double} shows the complete set of the elementary
blocks, together with their symbolic expressions, that are required
for the construction of the double scattering ladder and crossed
spectra. As seen from Fig.~\ref{fig:double},  the frequencies of
incoming and outgoing amplitudes are correlated, which is a direct
consequence of energy conservation during the scattering
processes \cite{wellens10,geiger10}.

\begin{figure}
\includegraphics[width=3.5cm]{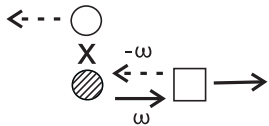}
\caption{An example of a double scattering diagram contributing to
the elastic crossed spectrum. The spectral response of the left atom
is elastic, and constructed as a product (denoted by the `X') of the
block which is the complex conjugate of the one shown in
Fig.~\ref{fig:double}(a), and of the block Fig.~\ref{fig:double}(c)
(one should bear in mind that, for circles, the direction of the
outgoing arrow is immaterial for the definition of the spectral
response \cite{shatokhin12a}). The spectral response of the right
atom  is represented by the block in Fig.~\ref{fig:double}(f). There
is no hatching of the square since, as discussed in the text, the
outgoing arrow that represents the detected field is elastic with
respect to the laser frequency. The overall mathematical expression
for this double scattering process is given in
Eq.~(\ref{eq:fig:eq}).} \label{fig:examp_cross}
\end{figure}

Furthermore, it should be mentioned that, for each of the elementary
blocks, a replacement of solid arrows with dashed ones and vice
versa yields complex conjugated blocks. Therefore, knowledge of the
spectral responses shown in Fig.~\ref{fig:double} suffices to obtain
an arbitrary single-atom spectral response needed to infer the
double scattering signal. Circles with one outgoing solid arrow [see
Figs.~\ref{fig:double}(a)-(d)] provide graphical representations of
the perturbative corrections of zeroth (no incoming arrows), first
(one incoming solid or dashed arrow), and second order (one dashed
and one solid incoming arrows) to the expectation value of the atomic dipole lowering operator $\la\s^-\ra$. Squares with two outgoing arrows [see
Figs.~\ref{fig:double}(e)-(h)] correspond to the perturbative
solutions for the inelastic component of the frequency correlation
function  (\ref{eq:spec_corr_func}). We put labels above the
arrows to denote the detunings of the corresponding waves from the
laser field frequency; in case of exact resonance, the labels are
omitted for brevity. Furthermore, we leave a shape blank if its
outgoing arrow is elastic with respect to the laser frequency (in
case of the squares, this rule applies to the arrow that is directed
towards the detector if the detected field is elastic with respect
to the laser frequency, see e.g. Fig.~\ref{fig:examp_cross}).
Otherwise, the shape is hatched. The expressions for the correlation
functions on the right hand sides of each of the diagrams in
Fig.~\ref{fig:double} can be found in \cite{shatokhin12a}.

To construct double scattering processes contributing to the ladder or crossed signals, one decomposes the total spectral response of each of the two atoms into its elastic and inelastic components, using the elementary building blocks from Fig.~\ref{fig:double} and their complex conjugates \cite{shatokhin12a}. Then the diagrammatic expansions for both atoms are reconnected self-consistently, using a set of rules \cite{shatokhin12a}, to form double scattering diagrams of either ladder [see Fig.~\ref{fig:processes}(a)] or crossed [see Fig.~\ref{fig:processes}(b)] types. We present an example of a double scattering diagram contributing to the elastic crossed spectrum in Fig.~\ref{fig:examp_cross}. Applying the rules of self-consistent combination of the building blocks to the relevant spectral response functions (see Fig.~\ref{fig:double}), we obtain
\be
{\rm  Fig}.~\ref{fig:examp_cross}=|\bar{g}|^2\int_{-\infty}^{\infty}\frac{d\omega}{2\pi}
\la\s^+\ra^{(0)}\la\s^-(-\omega)\ra^{(+)}P^{(-)}(\omega,0),
\label{eq:fig:eq}
\e
where $\bar{g}\propto (k_L\ell)^{-1}$, with $\ell$ the average interatomic distance. Detailed
expressions and numerous examples of the double scattering elastic and inelastic spectra can be found in \cite{geiger09}.

Concluding this section, we would like to mention that the
elementary single-atom blocks have a physical interpretation as 
effective nonlinear susceptibilities \cite{shatokhin12a}, which
describe the response of the laser-driven atom to weak probe fields
\cite{boyd}. Recently, it has been shown that there is a systematic
way of obtaining analytical expressions for such blocks in case of
an arbitrary number of probe fields \cite{shatokhin12b}. In future
work, these expressions will be incorporated into the theory of
nonlinear transport by classical scatterers \cite{wellens08} to
describe CBS of intense laser light in cold atomic gases
\cite{binninger12}.

\section{Generalization of the approach to atoms with degenerate dipole transitions}
\label{sec:vector}

\subsection{Fundamental double scattering processes}
\label{sec:fundam_vector}

The diagrammatic pump-probe approach, presented in the previous
section, ignores the polarization degree of freedom of the light.
This is closely related to the fact that, in
Sec.~\ref{sec:two-level}, we reduced the internal quantum structure
of the atomic dipole transitions to one ground and one excited
level. However, Sr or Rb atoms, studied in real experiments on CBS
of light, possess degenerate dipole transitions, what renders this
effect sensitive to the choice of the incoming and outgoing fields'
polarizations \cite{labeyrie03,kupriyanov06}. Our main goal will be
to include the internal degeneracy, as well as the vector character
of the electromagnetic field, into the diagrammatic pump-probe
approach. As in the case of scalar atoms, we will ensure, whenever
possible, a close correspondence with the results of the master
equation approach. Presently, such results have been made available
for double scattering from two Sr atoms
\cite{shatokhin05,shatokhin07,ralf11}.
\begin{figure}
\includegraphics[width=8cm]{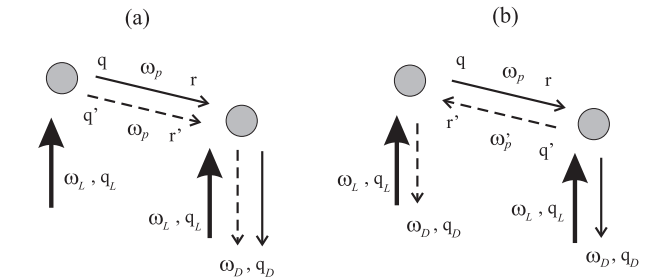}
\caption{Double scattering diagrams which survive the disorder
average -- for the case of vector electric fields, and atoms with
degenerate dipole transitions: (a)  ladder spectrum; (b) crossed
spectrum. The meaning of the labels $\omega_L$, $\omega_p$,
$\omega_p^\prime$, and $\omega_D$ is the same as in
Fig.~\ref{fig:processes}. Indices $q_L$, $q$, $q^\prime$,
 $r$, $r^\prime$, and $q_D$
refer to the polarization indices of the corresponding arrows in the
spherical basis (see text for further details).} \label{fig:model1}
\end{figure}

Inclusion of polarization and electronic degeneracy amounts
to a certain technical overhead, without affecting the basic
idea of the approach. Namely, the matrix dimension of the linear
system generalizing Eq.~(\ref{eq:MOBE}) will increase according to
the number of sublevels of the electronic ground and excited
states. Furthermore, the explicit form of the single-atom building
blocks will now depend on the choices of the pump, probe, and
detected fields' polarizations. However, our justification of the
classical description of the exchanged amplitudes, as presented in
Sec.~\ref{sec:two-level}, certainly remains true also for polarized
electric fields.

To begin with, let us consider a vectorial generalization of
the fundamental scattering processes (see Fig.~\ref{fig:processes}) that survive the disorder average in Fig.~\ref{fig:model1}.
Co-propagating inelastic
scattering amplitudes contribute to the ladder spectrum [see
Fig.~\ref{fig:model1}(a)],  and counter-propagating amplitudes
contribute to the crossed spectrum [see Fig.~\ref{fig:model1}(b)].
 In addition to the elements that are present in Fig.~\ref{fig:processes}, each of the arrows in
 Fig.~\ref{fig:model1} is now garnished by polarization indices. Unless otherwise stated, any such index $q$ corresponds to a unit polarization vector $\hat{\bf
 e}_q$ in the spherical basis:
\be \hat{\bf e}_{\pm 1}=\mp \frac{1}{\sqrt{2}}\lt(\hat{\bf e}_x\pm
i\hat{\bf e}_y\rt), \quad \hat{\bf e}_0=\hat{\bf e}_z, \e where
$\hat{\bf e}_x$, $\hat{\bf e}_y$, and $\hat{\bf e}_z$ are the unit
vectors in the Cartesian basis.

In general, arrows corresponding to the scattered fields carry a
pair of polarization indices. However, we will study the
CBS signal in exact backscattering direction, that
is, along the quantization axis set by the direction of the laser
wave. Therefore, the polarization of the backscattered field, alike
the laser field, can be specified by a single index.
\begin{figure}
\includegraphics[width=8cm]{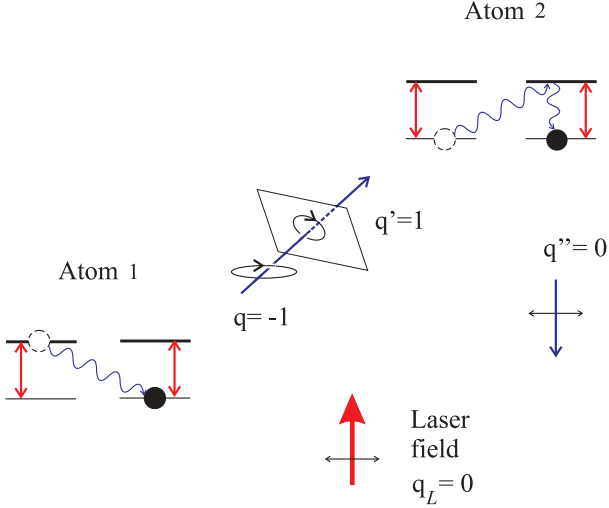}
\caption{(Color online) A linearly polarized laser wave (thick red
arrow) excites the $\pi$-transitions (double red thin arrows) of two
atoms with ground and excited state angular momenta $J_g$ and $J_e$
equal to $1/2$. Atom 1 emits a $\s_-$ polarized amplitude ($q=-1$)
towards atom 2, and, after projection of the polarization vector
onto the plane perpendicular to the line connecting both atoms,
excites a $\s_+$-transition thereof ($q'=+1$). Finally, a
$\pi$-polarized amplitude ($q''=0$) is emitted by atom 2 towards the
detector.} \label{fig:model2}
\end{figure}
Introducing a pair of polarization indices for the intermediate
arrows can be motivated with the aid of Fig.~\ref{fig:model2}, which
presents an example of a double scattering process of a linearly
polarized positive-frequency laser wave by two atoms with equal
angular momenta of the ground and excited states: $J_g=J_e=1/2$.
As evident from Fig.~\ref{fig:model2}, the polarizations of the waves emitted by
atom 1 ($q=-1$) and absorbed by atom 2 ($q^\prime=+1$) can be
different, hence the two indices for the intermediate amplitudes. For the positive-frequency wave, the probability
amplitudes of various combinations of $q, q^\prime$ are defined by
the projections thereof on the plane transverse to the line
connecting the atoms (see Fig.~\ref{fig:model2}), given by
$\overleftrightarrow{\boldsymbol\Delta}_{q^\prime q}\equiv \hat{\bf
e}^*_{q^\prime}\cdot\overleftrightarrow{\boldsymbol\Delta}\cdot\hat{\bf
e}_{q}$, with the projector on the transverse plane
$\overleftrightarrow{\boldsymbol\Delta}=\overleftrightarrow{\openone}-\hat{\bf
n}\hat{\bf n}$, and
\begin{align}
\overleftrightarrow{\openone}&=-\hat{\bf e}_{-1}\hat{\bf
e}_{+1}+\hat{\bf e}_{0}\hat{\bf e}_{0}-\hat{\bf e}_{+1}\hat{\bf
e}_{-1},\label{eq:identity}\\
\hat{\bf n}&=\frac{e^{i\phi}\sin\vartheta}{\sqrt{2}}\hat{\bf
e}_{-1}+\cos\vartheta\hat{\bf
e}_0-\frac{e^{-i\phi}\sin\vartheta}{\sqrt{2}}\hat{\bf e}_{+1}.
\label{eq:angles}
\end{align}
By analogy, it is easy to show that the complex conjugate amplitude of
the one shown in Fig.~\ref{fig:model2} is
proportional to
$(\overleftrightarrow{\boldsymbol\Delta}_{q^\prime q})^*=\hat{\bf
e}^*_q\cdot\overleftrightarrow{\boldsymbol\Delta}\cdot\hat{\bf
e}_{q^\prime}$.

It follows from the above that the double scattering processes shown
in Figs.~\ref{fig:model1}(a) and \ref{fig:model1}(b) are
proportional to the geometric factor $({\bf
e}_{r}^*\cdot\overleftrightarrow{\boldsymbol\Delta}\cdot\hat{\bf
e}_{q})({\bf
e}^*_{q^\prime}\cdot\overleftrightarrow{\boldsymbol\Delta}\cdot\hat{\bf
e}_{r^\prime})$, whose explicit form can
easily be obtained for arbitrary polarization indices using
Eqs.~(\ref{eq:identity}), (\ref{eq:angles}). Next, we need to
perform the configuration average over the random angles
$(\vartheta,\phi)$ -- which define the orientation of the vector
$\hat{\bf n}$ connecting the atoms with respect to the quantization
axis [see Eq.~(\ref{eq:angles})]. The resulting geometric weight for
diagrams in Fig.~\ref{fig:model1}(a),(b) reads \be \la
\overleftrightarrow{\boldsymbol\Delta}_{rq}\overleftrightarrow{\boldsymbol\Delta}_{q^\prime
r^\prime}\ra=\frac{1}{4\pi}\int_0^{\pi}\sin\vartheta
d\vartheta\int_0^{2\pi}\overleftrightarrow{\boldsymbol\Delta}_{rq}\overleftrightarrow{\boldsymbol\Delta}_{q^\prime
r^\prime}d\phi
. \label{eq:conf_aver} \e Finally, if
there are several polarization channels for double scattering, we
perform a summation over the corresponding polarization indices.

Each of the disorder-averaged geometric weights must be multiplied by the corresponding
double scattering spectral response, whose evaluation
from single-atom building blocks will be considered in the
subsequent sections.

\subsection{Diagrammatic expansion of the double scattering process}
\label{sec:expand} 
\begin{figure}
\includegraphics[width=8cm]{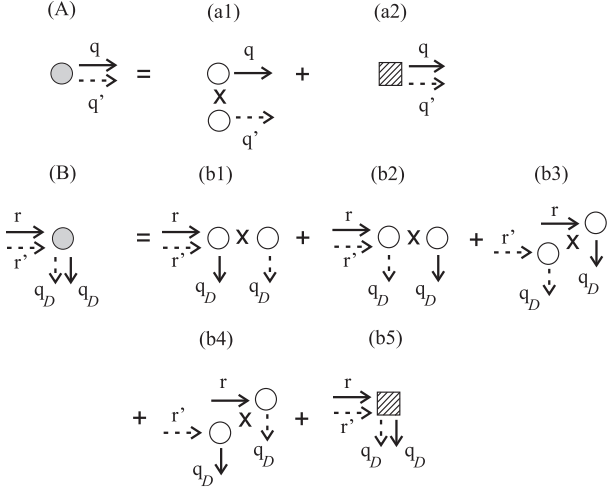}
\caption{Diagrammatic expansion of the double scattering process
depicted in Fig.~\ref{fig:model1}(a) into its elastic (blank circles) and inelastic
(hatched boxes) components. Left (A, B): Single-atom building blocks contributing to the
ladder spectrum. Right (a1-b5): Expansion of the single-atom building blocks
into elementary building blocks.} \label{fig:ladder_vector}
\end{figure}

\begin{figure}
\includegraphics[width=8cm]{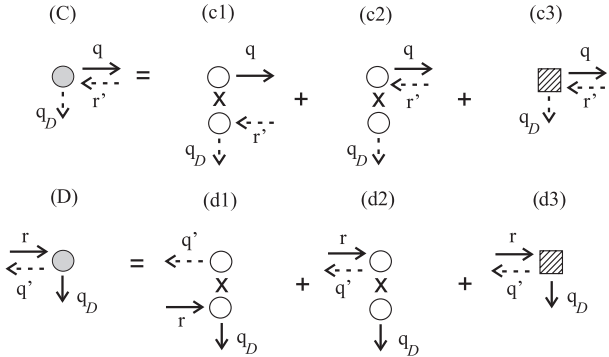}
\caption{Diagrammatic expansion of the double scattering process
depicted in Fig.~\ref{fig:model1}(b) into the elastic (blank circles) and inelastic
(hatched boxes) components. Left (C, D): Single-atom building blocks contributing to the
crossed spectrum. Right (c1-d3): Expansion of the single-atom building
blocks into elementary building blocks.} \label{fig:crossed_vector}
\end{figure}

After selecting the double scattering processes which
survive the disorder average, we proceed
by considering the two atoms and their incoming and outgoing classical fields in Fig.~\ref{fig:model1}(a),(b)
separately. In complete analogy with the case of scalar atoms
\cite{geiger10,wellens10,shatokhin12a}, the spectral response of either one of the atoms is split into an elastic and an inelastic component. It is convenient to represent these components graphically, as shown in
Figs.~\ref{fig:ladder_vector} and \ref{fig:crossed_vector}.

We remind that, to alleviate the diagrams, the arrows which represent the
laser field are not depicted in Figs.~\ref{fig:ladder_vector} and
\ref{fig:crossed_vector}. Note also that we do not yet assign the frequency
values to different arrows in Figs.~\ref{fig:ladder_vector},
\ref{fig:crossed_vector}: these will be determined in
the course of a self-consistent combination of the single-atom
responses into double scattering ladder and crossed spectral signals
(see Sec.~\ref{sec:self-consistent}). To facilitate establishing the
correspondence between the diagrams in
Figs.~\ref{fig:ladder_vector}, \ref{fig:crossed_vector} and \ref{fig:model1}, respectively, we depict the backscattered fields with
downward-directed arrows.

In each of these graphical equations, open circles with one outgoing
arrow and null, one or two incoming arrows describe the elementary
elastic building blocks. Circles are always combined in pairs by the symbols X. We will see below, in
Sec.~\ref{sec:elastic_building_blocks}, that pairs of circles
correspond to the factorized parts of the atomic dipole correlation
function, which describe the elastic spectral responses. The number of pairs of circles in the
graphical expansion of the building blocks
is equal to $2^n$, where $n$ is the number of incoming probe
fields \cite{shatokhin12a,shatokhin12b}.

Apart from the open circles, each of the graphical equations in
Figs.~\ref{fig:ladder_vector}, \ref{fig:crossed_vector} contains one
hatched square, with two outgoing arrows and null, one or two
incoming arrows. This corresponds to the inelastic elementary
building block, which can be derived from the fluctuating part of
the atomic dipole correlation function, see
Sec.~\ref{sec:inelastic1}.

Computation of the single-atom elementary elastic and inelastic
spectral responses is based on the formalism of the generalized OBE,
to be explained below.

\subsection{Generalized optical Bloch equations}
\label{sec:general_OBE} This section presents a step-by-step
generalization of the OBE formalism outlined in
Sec.~\ref{sec:two-level}, to the case of vector fields and atoms
with arbitrary dipole transitions. We set out by writing down the
expression for the classical bichromatic vector field \be {\bf
E}_{pp}(t)=\mathcal{E}\hat{\bf
e}_Le^{-i\omega_Lt}+\mathcal{E}^*\hat{\bf
e}^*_Le^{i\omega_Lt}+\epsilon\hat{\bf
e}_{r}e^{-i\omega_pt}+\epsilon^*\hat{\bf
e}^*_{r^\prime}e^{i\omega_pt}, \label{eq:vector_bi_EM} \e where the
meaning of $\mathcal{E}$, $\epsilon$, $\omega_L$, $\omega_p$ is the
same as in Eq.~(\ref{eq:1}), and $\hat{\bf e}_L$, $\hat{\bf e}_r$
($\hat{\bf e}^*_{r^\prime}$) are the unit polarization vectors of
the laser and probe fields, respectively.

To account for the vector nature of the atomic dipole transition, we
introduce vector raising and lowering atomic operators instead
of the operators $\s^+$ and $\s^-$. We will consider atoms
with total ground and excited state angular momenta $J_g$ and
$J_e$, respectively. Then the atomic raising and lowering operators,
${\bf D}^\dagger$ and ${\bf D}$, can be expressed using the
projection operators on the ground and excited state manifolds: \be
P_e=\sum_{m_e=-J_e}^{J_e}|J_em_e\ra\la J_em_e|,\;
P_g=\sum_{m_g=-J_g}^{J_g}|J_gm_g\ra\la J_gm_g|,
\label{eq:projectors}\e where $|J_em_e\ra$ ($|J_gm_g\ra$) denotes an
excited (ground) state sublevel with magnetic quantum number $m_e$
($m_g$). The raising and lowering parts of the atomic dipole
operator read \be {\bf D}^\dagger=\frac{1}{d}P_e\boldsymbol{\cal
D}P_g, \quad {\bf D}=({\bf D}^\dagger)^\dagger,
\label{eq:vector_raising}\e where $d\equiv \la J_e||{\cal
D}||J_g\ra$ is the reduced matrix element, and $\boldsymbol{\cal
D}=d({\bf D}^\dagger+{\bf D})$ is the atomic dipole moment operator.
Inserting the projectors (\ref{eq:projectors}) into
Eq.~(\ref{eq:vector_raising}), and using the Wigner-Eckart theorem
\cite{sakurai}, we obtain the following expression \be {\bf
D}^\dagger=\!\!\sum_{q=-1}^{1}\sum_{m_g=-J_g}^{J_g}\!\!\! \hat{\bf e}_q^*\la
J_g m_g,1 q|J_e m_g+q\ra|J_e m_g+q\ra\la J_g m_g|,
\label{eq:final_raising_vector}\e where $\la J_g m_g,1 q|J_e
m_g+q\ra$ denotes a Clebsch-Gordan coefficient, and one summation (over
$m_e$) was removed from Eq.~(\ref{eq:final_raising_vector}) owing to
the dipole transition selection rules \cite{sakurai}.

With the vector bichromatic field and dipole operators defined, we merely make the replacements
$E_{pp}(t)\rightarrow {\bf E}_{pp}(t)$, $\s^+\rightarrow {\bf D}^\dagger$,
$\s^-\rightarrow {\bf D}$ in Eq.~(\ref{eq:master}), to obtain its vector
generalization:
\begin{align}
\la\dot{Q}\ra&=\left\la\!-i\delta[{\bf D}^\dagger\!\cdot \!{\bf D},Q]\!
-\!\frac{i}{2}[\Omega ({\bf D}^\dagger\!\cdot \!\hat{\bf e}_L)\!+\!\Omega^*({\bf D}\!\cdot\! \hat{\bf e}_L^*),Q]\right.\n\\
&\left.+\gamma\lt({\bf D}^\dagger\cdot[Q,{\bf D}]+[{\bf D}^\dagger,Q]\cdot{\bf D}\rt)\right.\n\\
&\left. -\frac{i}{2}[g e^{-i\omega t}({\bf D}^\dagger\cdot \hat{\bf
e}_{r})+g^* e^{i\omega t}({\bf D}\cdot\hat{\bf
e}_{r^\prime}^*),Q]\right\ra. \label{eq:master_vector}
\end{align}
Finally, by choosing operators $Q$
from the complete set of operators (see Appendix \ref{app:projections}),  we translate the master equation
(\ref{eq:master_vector}) into a matrix equation for
the vector $\la{\bf Q}\ra$, in full analogy with the case of a
two-level atom [see Eq.~(\ref{eq:MOBE})]. Accordingly, we will refer
to the resulting system of equations,
\begin{align}
\la\dot{\bf Q}\ra&={\bf M}\la{\bf Q}\ra+{\bf
L}+ge^{-i\omega t}\boldsymbol{\Delta}_{r}^{(-)}\la{\bf Q}\ra\n\\
&+g^*e^{i\omega t}\boldsymbol{\Delta}_{r^\prime}^{(+)}\la{\bf Q}\ra,
\label{eq:gen_OBE}
\end{align}
as the {\it generalized} optical Bloch equations under bichromatic driving.

\subsection{Single-atom spectral correlation functions}
We characterize the spectral response of multilevel atoms in
different polarization channels by the tensor frequency correlation
function generalizing Eq.~(\ref{eq:spec_corr_func}): \begin{align}
I_{q^\prime q}(\nu^\prime,\nu)&=\frac{1}{(2\pi)^2}\int_{-\infty}^{\infty}dt\int_{-\infty}^{\infty}dt^\prime
e^{-it^\prime\nu^\prime+it\nu}\n\\
&\times \la D^\dagger_{q^\prime}(t^\prime)D_{q}(t)\ra,
\label{eq:corr_dip_vector}
\end{align}
where $D^\dagger_{q^\prime}(t^\prime)\equiv \hat{\bf e}_{q^\prime}\cdot {\bf D}^\dagger(t^\prime)$
and $D_{q}(t)\equiv \hat{\bf e}^*_{q}\cdot{\bf
D}(t)$. In complete analogy with the scalar case, we proceed
by decomposing the atomic dipole correlation function in
(\ref{eq:corr_dip_vector}) into a factorized and a fluctuating
part: \be \la D^\dagger_{q^\prime}(t^\prime)D_{q}(t)\ra=\la
D^\dagger_{q^\prime}(t^\prime) \ra\la D_{q}(t)\ra+\la \Delta
D^\dagger_{q^\prime}(t^\prime)\Delta
D_{q}(t)\ra,\label{eq:fact_and_fluct_vector}\e with
the fluctuating part $\Delta D_q$ defined in strict analogy to
Eq.~(\ref{eq:fact_and_fluct}). Both of the correlation functions on the right hand side of Eq.~(\ref{eq:fact_and_fluct_vector}) can be found by solving Eq.~(\ref{eq:gen_OBE}). Plugging the right hand side
of Eq.~(\ref{eq:fact_and_fluct_vector}) into
Eq.~(\ref{eq:corr_dip_vector}), we obtain a tensor correlation function which generalizes Eq.~(\ref{eq:int_el_in}): \be
I_{q^{\prime}q}(\nu^\prime,\nu)=I^{\rm
el}_{q^{\prime}q}(\nu^\prime,\nu)+I^{\rm
in}_{q^{\prime}q}(\nu^\prime,\nu). \label{eq:int_el_in_vector}\e
Once again, the elastic and inelastic components arise from the factorized
and fluctuating parts of the atomic dipole correlation function,
respectively. Apart from the frequencies $\nu$, $\nu^\prime$ and
polarization indices $q,q^\prime$, the correlation functions in Eq.~(\ref{eq:int_el_in_vector}) depend also on the frequency
$\omega$ and the polarization indices $r$, $r^\prime$ of the incoming wave. These dependencies
will be reflected in the diagrammatic representation of the single
atom blocks, to be introduced below.
\subsection{Building blocks}
\label{sec:building_blocks} The evaluation of the single-atom
building blocks in the vectorial case is again completely analogous
to the scalar one. To see this, it is important to realize that,
regardless of the structure of the dipole transition and the
polarization indices of the driving and scattered fields, the
spectral response functions, expanded to second order in the probe
field amplitude, satisfy the same energy conservation conditions as
their scalar analogs studied in detail in \cite{wellens10,geiger10}. Namely,
the response functions contain $\delta$-functions originating from integrations over
time in Eq.~(\ref{eq:corr_dip_vector}), under the assumption of
stationarity of the atomic dipole correlation functions $\la
D^\dagger_{q^\prime}(t^\prime)D_{q}(t)\ra=\la
D^\dagger_{q^\prime}(t^\prime-t)D_{q}(0)\ra$, what entails strict
relations between the incoming and outgoing frequencies.

\begin{figure}
\includegraphics[width=8.5cm]{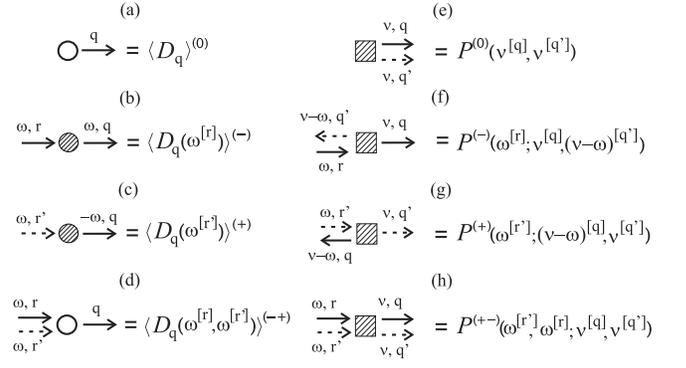}
\caption{Elementary single-atom building blocks, together with the
corresponding spectral response functions in the vector case.
(a)-(d) Complex scattering amplitudes associated with the
perturbative solutions of Eq.~(\ref{eq:gen_OBE}); (e)-(h) blocks
associated with the perturbative solutions for the fluctuating part
of the atomic dipole correlation function, see
Eq.~(\ref{eq:fact_and_fluct_vector}). The notation $\omega^{[q]}$ for
a positive-(negative-)frequency wave implies a wave with frequency
$\omega$ and polarization described by the unit vector $\hat{\bf e}_q$ ($\hat{\bf e}_q^*$). Blank and hatched shapes denote elastic and inelastic spectral responses, respectively, in full analogy to the scalar case, see Fig.~\ref{fig:double}.}
\label{fig:blocks_vector}
\end{figure}

We incorporate these relations into the diagrammatic
representation of the elementary single-atom building blocks (see Fig.~\ref{fig:blocks_vector}).
As seen from Fig.~\ref{fig:blocks_vector}, we introduce the
same type (elastic or inelastic) and number of the spectral response
functions as in the scalar case (see Fig.~\ref{fig:double}). In addition to the graphical elements that are already present in the scalar case, each incoming and outgoing arrow in Fig.~\ref{fig:blocks_vector} carries
a polarization index.

We will now explain how to find
explicit expressions for the vector spectral responses on the right hand
sides of the graphical equations in Fig.~\ref{fig:blocks_vector}.

\subsubsection{Elastic building blocks}
\label{sec:elastic_building_blocks} All the elastic spectral
response functions appearing on the right hand side in
Fig.~\ref{fig:blocks_vector}(a)-(d) can be obtained directly from
the stationary perturbative solutions of the generalized OBE~(\ref{eq:gen_OBE}), to second order in the probe field amplitude. Setting the left hand side of Eq.~(\ref{eq:gen_OBE}) to zero, it is easy to obtain the following perturbative solutions:
\beml
\begin{align}
\la{\bf Q}\ra^{(0)}&={\bf G}{\bf L},\label{eq:solution_gen_OBE_a}\\
\la{\bf Q}(\omega^{[r]})\ra^{(-)}&={\bf
G}(-i\omega)\boldsymbol{\Delta}^{(-)}_r\la{\bf Q}\ra^{(0)},\\
\la{\bf Q}(\omega^{[r^\prime]})\ra^{(+)}&={\bf
G}(i\omega)\boldsymbol{\Delta}^{(+)}_{r^\prime}\la{\bf Q}\ra^{(0)},\\
\la{\bf Q}(\omega^{[r]},\omega^{[r^\prime]})\ra^{(-+)}&={\bf
G}\boldsymbol{\Delta}^{(+)}_{r^\prime}\la{\bf Q}(\omega^{[r]})\ra^{(-)}\n\\
&+{\bf G}\boldsymbol{\Delta}^{(-)}_r\la{\bf
Q}(\omega^{[r^\prime]})\ra^{(+)},\label{eq:solution_gen_OBE_d}
\end{align}
\label{eq:solution_gen_OBE} \eml where \be {\bf
G}(z)=\frac{1}{z-{\bf M}} \e is the free propagator
governing the internal dynamics of the laser-driven, damped atom,
and ${\bf G}\equiv {\bf G}(0)$.

Using the perturbative solutions (\ref{eq:solution_gen_OBE}), the expressions
for the elementary blocks with one solid (dashed) outgoing arrow
carrying the index $q$ [see Figs.~\ref{fig:blocks_vector}(a)-(d)] can be expressed as scalar
products with the projection vectors ${\bf V}_q$ (${\bf U}_q$), respectively (see Appendix \ref{app:projections}). For example, the zeroth-order projections yield
\be
\la D_q\ra^{(0)}={\bf V}_q\cdot\la{\bf Q}\ra^{(0)},\quad \la D^{\dagger}_q\ra^{(0)}={\bf U}_q\cdot\la{\bf Q}\ra^{(0)}.
\e
The remaining elementary building blocks are constructed analogously.

\subsubsection{Inelastic building blocks}
\label{sec:inelastic1}
The starting point for the derivation of the inelastic building blocks is the introduction of the stationary vector correlation functions
\beml
\begin{align}
{\bf f}_q(\tau)&=\la \Delta{\bf Q}(\tau)\Delta D_q\ra,\label{eq:inel_blocks1}
\\
{\bf h}_{q^{\prime}}(\tau)&=\la \Delta D^{\dagger}_{q^{\prime}}\Delta{\bf Q}(\tau)\ra,
\end{align}
\label{eq:f_h}
\eml
where $\tau\geq 0$.

Application of the quantum regression theorem to Eq.
(\ref{eq:inel_blocks1}) leads to the following equation of motion
for the vector ${\bf f}_q$ [compare with Eq.~(\ref{eq:gen_OBE})]:
\begin{equation}
\dot{\bf f}_q={\bf M f}_q+ge^{-i\omega
t}\boldsymbol{\Delta}_r^{(-)}{\bf f}_q +g^*e^{i\omega
t}\boldsymbol{\Delta}_{r^\prime}^{(+)}{\bf f}_q,
\label{eq:inel_corr_vector}
\end{equation}
and the equation for  ${\bf h}_{q^{\prime}}$ is obtained upon
replacing ${\bf f}_q\rightarrow {\bf h}_{q^{\prime}}$. The temporal
evolutions of the vector functions ${\bf f}_q$ and ${\bf
h}_{q^{\prime}}$ are, of course, different from each other, due to the different
initial conditions, ${\bf f}_q(0)\neq {\bf h}_{q^{\prime}}(0)$, see
Eq.~(\ref{eq:f_h}). We solve Eq.~(\ref{eq:inel_corr_vector}) perturbatively using Laplace transform; the solutions for ${\bf h}_{q^{\prime}}$ follow by analogy. As we will see below, Laplace transforms of ${\bf f}_q$ and ${\bf h}_{q^{\prime}}$ define the outgoing negative-frequency amplitude with polarization $q^{\prime}$ and positive-frequency amplitude with polarization $q$, respectively, of the inelastic building blocks in Figs.~\ref{fig:blocks_vector}(e)-(h). We
have
\begin{widetext}
\beml
\begin{align}
\tilde{\bf f}^{(0)}_q(z^{\prime\prime})&={\bf G}(iz^{\prime\prime}){\bf f}^{(0)}_q(0),\\
\tilde{\bf f}^{(+)}_q(\omega^{[r^\prime]}; z^{\prime\prime})&={\bf G}(iz^{\prime\prime}+i\omega)\lt\{\boldsymbol{\Delta}^{(+)}_{r^\prime}\tilde{\bf f}^{(0)}_q(z^{\prime\prime})+{\bf f}^{(+)}_q(\omega^{[r^\prime]}; 0)\rt\},\\
\tilde{\bf f}^{(-)}_q(\omega^{[r]}; z^{\prime\prime})&={\bf G}(iz^{\prime\prime}-i\omega)\lt\{\boldsymbol{\Delta}^{(-)}_r\tilde{\bf f}^{(0)}_q(z^{\prime\prime})+{\bf f}^{(-)}_q(\omega^{[r]}; 0)\rt\},\\
\tilde{\bf f}^{(+-)}_q(\omega^{[r^\prime]},\omega^{[r]}; z^{\prime\prime})&={\bf
G}(iz^{\prime\prime})\lt\{\boldsymbol{\Delta}^{(-)}_r\tilde{\bf
f}^{(+)}_q(\omega^{[r^\prime]};
z^{\prime\prime})+\boldsymbol{\Delta}^{(+)}_{r^\prime}\tilde{\bf
f}^{(-)}_q(\omega^{[r]}; z^{\prime\prime}) +{\bf
f}^{(+-)}_q(\omega^{[r^\prime]},\omega^{[r]}; 0)\rt\},
\end{align}
\label{eq:solf}
\eml
\end{widetext}
where $z^{\prime\prime}={\rm Im}(z)$, and the vectors of the initial conditions ${\bf f}^{(0)}_q(0)$,
${\bf f}^{(+)}_q(\omega^{[r^\prime]}; 0)$,  ${\bf
f}^{(-)}_q(\omega^{[r]}; 0)$ and ${\bf
f}^{(+-)}_q(\omega^{[r^\prime]},\omega^{[r]}; 0)$ are given in
Appendix \ref{app:initial}. Now, the outgoing positive- and
negative-frequency fields of the inelastic building blocks follow
via scalar products of the obtained perturbative solutions with the
projection vectors ${\bf V}_q$ and ${\bf U}_{q^\prime}$,
respectively, yielding the following expressions for the inelastic
building blocks:
\begin{widetext}
\beml
\begin{align}
P^{(0)}(\nu^{[q]},\nu^{[q^\prime]})&={\bf U}_{q^\prime}\cdot\tilde{\bf f}^{(0)}_q(\nu)+{\bf V}_q\cdot\tilde{\bf h}^{(0)}_{q^\prime}(-\nu),\\
P^{(-)}(\omega^{[r]};\nu^{[q]},(\nu-\omega)^{[q^\prime]})&=
{\bf U}_{q^\prime}\cdot\tilde{\bf f}^{(-)}_q(\omega^{[r]}; \nu)+{\bf V}_q\cdot\tilde{\bf h}^{(-)}_{q^\prime}(\omega^{[r]}; \omega-\nu),\\
P^{(+)}(\omega^{[r^\prime]};(\nu-\omega)^{[q]},\nu^{[q^\prime]})&=
{\bf U}_{q^\prime}\cdot\tilde{\bf f}^{(+)}_q(\omega^{[r^\prime]}; \nu-\omega)+{\bf V}_q\cdot\tilde{\bf h}^{(+)}_{q^\prime}(\omega^{[r^\prime]}; -\nu),\\
P^{(+-)}(\omega^{[r^\prime]},\omega^{[r]};\nu^{[q]},\nu^{[q^\prime]})&=
{\bf U}_{q^\prime}\cdot\tilde{\bf
f}^{(+-)}_q(\omega^{[r^\prime]},\omega^{[r]}; \nu)+{\bf
V}_q\cdot\tilde{\bf
h}^{(+-)}_{q^\prime}(\omega^{[r^\prime]},\omega^{[r]}; -\nu),
\end{align}
\label{eq:inel_blocks}
\eml
\end{widetext}
with the values of $z$ in every expression above fixed by the energy conservation relation, in strict analogy with the scalar case \cite{geiger10}.

\subsection{Self-consistent combination of single-atom building blocks}
\label{sec:self-consistent} In the previous section we defined the
elementary single-atom building blocks. Now, we will discuss the
rules of their self-consistent combination into double-scattering
contributions to CBS. For non-degenerate dipole transitions, these rules were
elaborated in \cite{shatokhin12a,shatokhin12b}.

As already mentioned in Sec.~\ref{sec:building_blocks}, for fixed values of the polarization
indices, the number of the elementary
elastic and inelastic response functions is the same as in the scalar case. Furthermore, these response functions exhibit the same relations between the frequencies of the incoming and outgoing fields. Therefore, the rules of the self-consistent combination that were formulated for
non-degenerate dipole transitions are valid also in the present case.

To be self-contained, we here briefly remind the reader of how to
construct the double scattering signal using single-atom
responses. To obtain the ladder spectrum, we connect the outgoing
arrows of each of the diagrams on the right hand side of the
graphical equation (A) with the incoming arrows of those of
equation (B) in Fig.~\ref{fig:ladder_vector}, respecting the
direction and character (solid or dashed) of the arrows. The
frequency values of all the arrows are assigned according to the
definitions of the elementary single-atom building blocks given in
Fig.~\ref{fig:blocks_vector}. If the frequency of an intermediate
arrow that is distinct from the laser frequency changes its value
upon the scattering process, it is integrated over. Finally, the two
downward arrows corresponding to the backscattered signal in a given
polarization channel should bear the same polarization indices and
frequency values (equal to $\nu$ for the inelastic component).
Application of these rules to the diagrammatic expansions (A) and
(B) in Fig.~\ref{fig:ladder_vector} results in six contributions --
(a1)(b1), (a1)(b2), (a1)(b3), (a1)(b4), (a2)(b1), (a2)(b2) -- to the
elastic, and four contributions -- (a1)(b5), (a2)(b3), (a2)(b4),
(a2)(b5) -- to the inelastic component of the double scattering
ladder spectrum. For example, the combination of diagrams (a2) and
(b5) in Fig.~\ref{fig:ladder_vector} yields the result
\begin{align}
\label{eq:Ladder_spec}
{\rm (a2)(b5)}&=|\bar g|^2\la \overleftrightarrow{\boldsymbol{\Delta}}_{q r}\overleftrightarrow{\boldsymbol{\Delta}}_{r^\prime q^\prime}\ra\int_{\infty}^{\infty}\frac{d\omega}{2\pi}P^{(0)}(\omega^{[q]},\omega^{[q^\prime]})\n\\
&\times
P^{(+-)}(\omega^{[r^\prime]},\omega^{[r]};\nu^{[q_D]},\nu^{[q_D]}).
\end{align}
Using this example, it is easy to construct the expressions for other contributions by analogy.

To obtain the crossed signal, we apply the same rules to the
graphical equations (C) and (D) in Fig.~\ref{fig:crossed_vector}.
Here, a subtlety arises when combining diagrams (c2) and (d2). Such a combination is forbidden since it features a closed loop including two amplitudes cycling
between the two circles without an outgoing amplitude \cite{wellens08,shatokhin12a,shatokhin12b}. Excluding the forbidden diagram, we obtain five contributions
-- (c1)(d1), (c1)(d2), (c2)(d1), (c2)(d3), and (c3)(d2) to the elastic, and three contributions -- (c1)(d3), (c3)(d1), and (c3)(d3) -- to the inelastic spectrum of CBS.

Finally, after summation over the relevant values of the intermediate polarization indices
$q$, $q^\prime$, $r$, and $r^\prime$, one obtains the result for the double scattering ladder and crossed spectra in a given polarization channel.

\section{Application: Double scattering by optically pumped atoms}
\label{sec:application}

\subsection{Formulation of the problem}
\label{sec:formulation_problem} In this section we apply the
formalism developed in Sec.~\ref{sec:vector} to calculate the double
scattering signal from {\em optically pumped} atoms in the helicity
preserving (h $\parallel$ h) polarization channel. This scenario is very different from the one where multiple scattering of a weak laser field
from degenerate atoms in the thermal equilibrium state was considered \cite{labeyrie99,labeyrie03,jonckheere00,mueller02,kupriyanov03}.

It is known that laser light with arbitrary polarization causes optical pumping
\cite{grynberg10}, that is, a non-equilibrium redistribution of the atomic ground state's sublevels' populations. The simplest situation
arises in the case of a circularly polarized laser field (for definiteness, we assume $\s_+$-polarization): Such a
field pumps the atoms into a transition with the maximal ground state
magnetic quantum number $m_g=J_g$. For the excited state angular
momenta $J_e=J_g-1$ and $J_e=J_g$ such a state is ``dark'', in the sense that the atoms get transparent for the laser light
\cite{gao93}. The only nontrivial situation leading to a CBS
signal corresponds to the transition $J_g\rightarrow J_e=J_g+1$. Therefore, henceforth we will exclusively deal with the case $J_e\equiv J_g+1$.
\begin{figure}
\includegraphics[width=8cm]{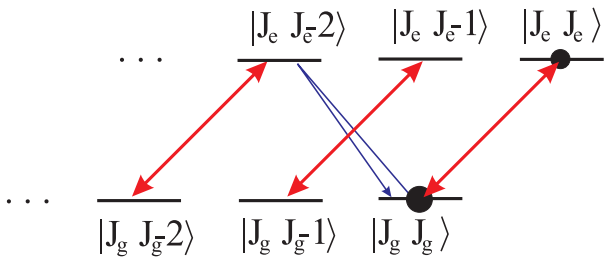}
\caption{(Color online) Degenerate dipole transition $J_g\rightarrow
J_e=J_g+1$ driven by $\s_+$-polarized light (thick red double
arrows). In the notation of the ground and excited state sublevels,
the first number refers to the angular momentum and the second one
to the magnetic quantum number. The black dots show that the
populations in the steady-state limit are distributed among the
states $|J_g\,J_g\ra$ and $|J_e\,J_e\ra$, with $J_e = J_g + 1$. The CBS signal in the helicity
preserving channel appears due to the double scattering process
on the transition between $|J_g\,J_g\ra$ and $|J_e\,J_e\!-\!2\ra$ depicted by the thin blue arrow.}
\label{fig:helicity_preserving}
\end{figure}

In Fig.~\ref{fig:helicity_preserving} we present a steady state
population distribution for an atom with such a transition optically
pumped by a $\s_+$-polarized laser field. Apart from that, in
Fig.~\ref{fig:helicity_preserving}, we depict a scattering process
which leads to a signal in the h $\parallel$ h polarization channel.
This (double) scattering process is mediated by the excited state
sublevel $|J_e\,J_e\!-\!2\ra$ (we remind the reader that first
and second symbol refer to the total angular momentum and to magnetic
quantum number, respectively). In the linear scattering regime, the
relevant levels are the three sublevels $|J_e\,J_e\!-\!2\ra$,
$|J_g\,J_g\ra$, and $|J_e\,J_e\ra$. Therefore, for any ground state
angular momentum, the ground state degeneracy becomes immaterial,
and perfect phase coherence of the CBS signal is predicted
\cite{kupriyanov04}. Does this imply that, in the nonlinear
scattering regime, the enhancement factor decays in the same way as
it does for $J_g=0$ as a function of the laser field strength? As
evident from Fig.~\ref{fig:helicity_preserving}, when two or more
laser photons are involved in the scattering process, the state
$|J_e\,J_e\!-\!2\ra$ can be coupled to the ground state sublevel
$|J_g\,J_g\!-\!2\ra$ (if $J_g\geq 1$), such that the atom
effectively becomes an N-type four-level system with the ground
state sublevels $|J_g \,J_g\!-\!2\ra$, $|J_g\, J_g\ra$ and the
excited state sublevels $|J_e\,J_e\!-\!2\ra$, $|J_e\,J_e\ra$. In
this case, the ground state degeneracy does come into play even
though the atoms are optically pumped. Below, we will explore the
effect of the internal degeneracy in optically pumped atoms
quantitatively, using the diagrammatic pump-probe approach.

\subsection{Selection of the polarization indices}
\label{sec:polarization} The qualitative consideration of
Sec.~\ref{sec:formulation_problem} allows us to identify all the
polarization indices of the single-atom blocks in
Figs.~\ref{fig:ladder_vector} and \ref{fig:crossed_vector}. We
recall that the indices $r$, $r^\prime$ describe the incoming waves,
and $q$, $q^\prime$ the outgoing ones; the index $q_D$ corresponds
to the polarization of the detected signal.

Let us first consider the ladder contribution, see
Fig.~\ref{fig:ladder_vector}. It is easy to see that
$q=q^\prime=+1$, since this corresponds to the polarization of the
field radiated by an atom that is optically pumped by a
$\s_+$-polarized laser field. Indices $r$, $r^\prime$ correspond to
the $\s_-$-polarized probe field depicted by the thin blue arrow in
Fig.~\ref{fig:helicity_preserving}, hence, $r=r^\prime=-1$. Finally,
detection in the parallel helicity channel means that $q_D=-1$. As
regards the crossed contribution, see Fig.~\ref{fig:crossed_vector},
we likewise obtain, for diagram (C): $r^\prime=q_D=-1$, $q=+1$, and
for diagram (D): $r=q_D=-1$, $q^\prime=+1$.

It follows from the above that, both, the ladder and crossed
contributions are proportional to the geometric weight
$\la|\overleftrightarrow{\boldsymbol{\Delta}}_{-1,+1}|^2\ra$, for
any $J_g$. Using the definitions (\ref{eq:identity}),
(\ref{eq:angles}), and (\ref{eq:conf_aver}), we easily perform the
angular integrations to obtain
$\la|\overleftrightarrow{\boldsymbol{\Delta}}_{-1,+1}|^2\ra=2/15$.

\subsection{Some basic properties of the building blocks for optically pumped atoms}
\label{sec:blocks_pumped} With the polarization indices fixed, the
elementary single-atom building blocks required for the evaluation
of the double scattering signal in the h $\parallel$ h polarization
channel can readily be evaluated using
Eqs.~(\ref{eq:solution_gen_OBE}) and (\ref{eq:inel_blocks}). Some of
these elementary blocks vanish identically in this channel, what
reduces the total number of the double scattering diagrams. First,
let us consider the elementary block shown in
Fig.~\ref{fig:blocks_vector}(a) (or its complex conjugate) with the
{\it downward}-directed arrow. Indeed, the corresponding amplitude
describes single scattering and must have the same polarization as
the laser field. Its contribution therefore vanishes in the h
$\parallel$ h polarization channel (where $q_D=-1$, as opposed to
$q=q^\prime=+1$ for the incident laser). By the same argument, all
double scattering diagrams containing the blocks (b1), (b2) (Fig.
\ref{fig:ladder_vector}), (c2) and (d2) (Fig.
\ref{fig:crossed_vector}) yield zero contribution. Second, let us
examine the block (b4) in Fig.~\ref{fig:ladder_vector} which is
composed of the two elementary blocks (see
Fig.~\ref{fig:blocks_vector}(c)) describing phase conjugation
processes of the probe fields in the presence of the laser field
\cite{shatokhin12a}, whereupon the incoming solid arrow turns into
the outgoing dashed arrow and vice versa. These are nonlinear
transformations of the probe fields which can only take place if the
probe and laser field polarizations coincide. But this is not the
case in the helicity-preserving channel (where $r=r^\prime=-1$ and
$q=q^\prime=+1$, see Sec.~\ref{sec:polarization}), hence, there is
no contribution to the ladder spectrum due to the block (b4).

After excluding the diagrams that do not contribute in the h
$\parallel$ h polarization channel, we end up with four double
scattering diagrams contributing each to the ladder and to the
crossed spectrum. We now consider the elastic and inelastic
components of both spectra separately.

\subsection{Elastic component}
\label{sec:elastic_component} The elastic ladder and crossed double
scattering spectra are obtained by combining diagrams (a1) and (b3)
in Fig.~\ref{fig:ladder_vector} and diagrams (c1) and (d1) in
Fig.~\ref{fig:crossed_vector}, respectively. It is evident that the
resulting ladder and crossed diagrams contain the same elementary
blocks. Hence, as expected \cite{kupriyanov04}, the elastic
component of the double scattering contribution to CBS yields
perfect interference contrast in the parallel helicity channel. We
have phenomenologically deduced an analytical expression for these
intensities which, as we have checked, exactly coincides with the result based on the numerical solution of
the OBE (see above) for arbitrary choice of the parameters $\Omega$, $\delta$, and $J_g$: \be L_{\rm el}=C_{\rm el}=\frac{1}{(4
J_g+1)^2}\frac{1}{1+(\delta/\gamma)^2}\frac{s}{(1+s)^4},
\label{eq:analytic_elastic} \e where we dropped a common geometric
prefactor, and introduced the saturation parameter \be
s=\frac{1}{2}\frac{\Omega^2}{\gamma^2+\delta^2}.
\label{eq:saturation_parameter}\e For $J_g=0$,
Eq.~(\ref{eq:analytic_elastic}) reduces to the result for Sr atoms
derived using the master equation approach \cite{shatokhin05}.

As already noted, perfect interference contrast, following from Eq.~(\ref{eq:analytic_elastic}), is a consequence of the optical pumping, whereupon the internal degeneracy does not play any role. In the opposite case of degenerate atoms in the thermal equilibrium (all ground state sublevels are equally populated), the contrast is in general $<1$, but its maximum value is restored in the semiclassical limit $J_g\to \infty$ \cite{cord_thesis}.

\subsection{Inelastic spectrum}
\label{sec:inelastic} The sum of the remaining self-consistent
combinations of diagrams: (a1)(b5) + (a2)(b3) + (a2)(b5) (see
Fig.~\ref{fig:ladder_vector}), yields inelastic ladder, and the sum
(c1)(d3) + (c3)(d1) + (c3)(d3) (see Fig.~\ref{fig:crossed_vector})
-- inelastic crossed spectra. Below, we present our numerical
results obtained by substituting solutions of Eqs.
(\ref{eq:solution_gen_OBE}) and (\ref{eq:inel_blocks}) into the
above graphical equations, along with a qualitative discussion of
how the internal degeneracy of optically pumped atoms affects the
inelastic CBS spectra.

In the inelastic scattering regime, the laser field couples the
excited state of the CBS transition to the unpopulated ground state
sublevel with $m_g=J_g-2$ (see Fig.~\ref{fig:helicity_preserving}).
Since such a coupling is impossible for atoms with $J_g=0$ and
$J_g=1/2$, these two types of transitions are expected to exhibit
similar behavior in the helicity preserving channel. And indeed, as
our calculations show (Figs. \ref{fig:weakly}(a), (b)  and
Fig.~\ref{fig:spectra_Sr}), the inelastic spectra for $J_g=1/2$
coincide, up to a prefactor $1/9=(4 J_g+1)^{-2}$, with the double
scattering spectra for the transition with $J_g=0$. Since the same
prefactor appears in the expression for the elastic intensities, see
Eq.~(\ref{eq:analytic_elastic}), the enhancement factors must
coincide for the transitions with $J_g=1/2$ and $J_g=0$, for
arbitrary parameters of the laser field.

For atoms
with $J_g\geq 1$, the CBS transition shares a common excited state
with the laser-driven transition  $|J_g\;J_g\!-\!2\ra\leftrightarrow|J_e\;J_e\!-\!2\ra$ (see
Fig.~\ref{fig:helicity_preserving}), what leads to qualitatively different spectra in the weakly
and strongly inelastic scattering regimes, as compared to the case of the non-degenerate atoms.

Below, we illustrate the above claims with numerical results for
different values of $J_g$.

\subsubsection{Weakly inelastic scattering}
\label{sec:weak_inelastic}
\begin{figure}
\includegraphics[width=8cm]{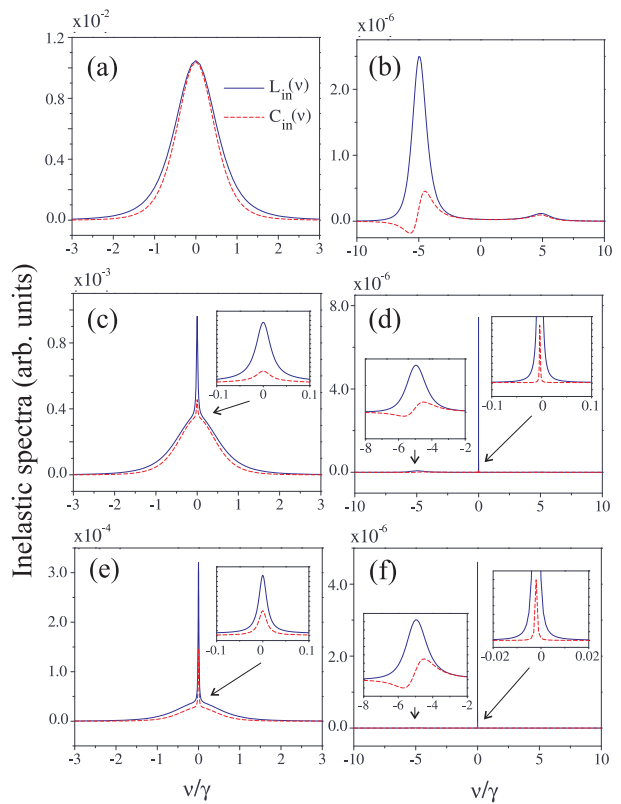}
\caption{(color online) Inelastic ladder (solid) and crossed
(dashed) double scattering CBS spectra in the weakly inelastic
regime ($\Omega=0.3\gamma$), for four different dipole transitions.
Top: $J_g=0$ and $J_g=1/2$. Both spectra coincide after rescaling
the $J_g=1/2$ plots by the factor 1/9.  Middle: $J_g=1$, bottom:
$J_g=3$. Left column: Exact resonance, $\delta=0$; right column:
Detuned driving, $\delta=5\gamma$. Insets magnify narrow resonances
that emerge for $J_g \geq 1$, centered at the driving frequency for
$\delta=0$, and slightly shifted towards a more pronounced sideband
for $\delta\neq 0$. In cases (d) and (f), sidebands at
$\delta=5\gamma$ exist, but are not resolved on this scale.}
\label{fig:weakly}
\end{figure}
Figure \ref{fig:weakly} shows several examples of the spectra for
the case $\Omega=0.3\gamma$. By virtue of
Eq.~(\ref{eq:saturation_parameter}), this corresponds to the weakly
inelastic regime, $s\ll 1$, for arbitrary detunings $\delta$. The
results for the transitions $J_g=0$ and $J_g=1/2$ coalesce in
Fig.~\ref{fig:weakly}(a) and (b) after rescaling the $J_g= 1/2$
signal with the prefactor $(4 J_g +1)^2$, see our discussion above.
In the resonant case (left panels), ladder and crossed spectra
exhibit an inelastic Rayleigh peak with a width of the order of
$\gamma$, centered at $\nu=0$; in the detuned case (right panels),
both spectra contain two sidebands centered at $\nu=\pm \delta$. The
detailed analytical and numerical results for double scattering
spectra and a physical interpretation thereof were presented for the
case of Sr atoms ($J_g=0$) in \cite{ralf11,shatokhin07}. We stress that,
here and in Sec.~\ref{sec:saturated} below, the double scattering
CBS spectra for Sr atoms calculated using the master
equation approach \cite{shatokhin07}  coincide with the ones found within the diagrammatic pump-probe approach \cite{ralf11}.

Starting from $J_g=1$, both the ladder and crossed spectra exhibit,
in addition to the spectrally wide features that are present in the
case of $J_g=0$ and $J_g=1/2$, narrow resonances centered at the
laser frequency, $\nu=0$ [in the detuned case, the position of the narrow resonance is slightly shifted towards a more pronounced sideband,
see Figs.~\ref{fig:weakly}(d), (f)].

Subnatural linewidth resonances are
typical for atoms with degenerate dipole transitions
\cite{grison91,javanainen92,gao94a,berman08,polder76,gao94b}.
Using the insights gained in these previous works, the emerging narrow peaks in the double scattering CBS spectra in the case $J_g\geq 1$ can
straightforwardly be explained \cite{berman08}:
Namely, since the system is optically pumped, an additional time scale, the finite life
time of unpopulated magnetic ground state sublevels, emerges. Associated
with this lifetime, optically pumped atoms acquire an effective subnatural width $\approx s\gamma$.
This width shows up in the CBS spectra as an additional narrow peak centered near $\nu=0$, when a field scattered
from another atom couples to the unpopulated ground state sublevel via a laser field (see Fig.~\ref{fig:helicity_preserving}). 

We will see below in Sec.~\ref{sec:saturated} that, in the double scattering spectral signal from atoms with $J_g\gg 1$, ultranarrow peaks can appear even in the strong saturation regime, $s\gg 1$. In that case, the physical origin of the narrow resonances is quantum interference between stimulated emission processes.

\subsubsection{Inelastic scattering from saturated atoms}
\label{sec:saturated}
Since saturation sets in for  $s\gtrsim 1$, the {\it narrow} features in the CBS spectra
then disappear (unless $J_g\gg 1$).
The degeneracy of the Zeeman sublevels is lifted by the dynamic Stark effect, and the
shape of
the double scattering CBS spectra can be understood by analyzing the dressed state structure of the
relevant dipole transitions of the atoms.
\begin{figure}
\includegraphics[width=6.5cm]{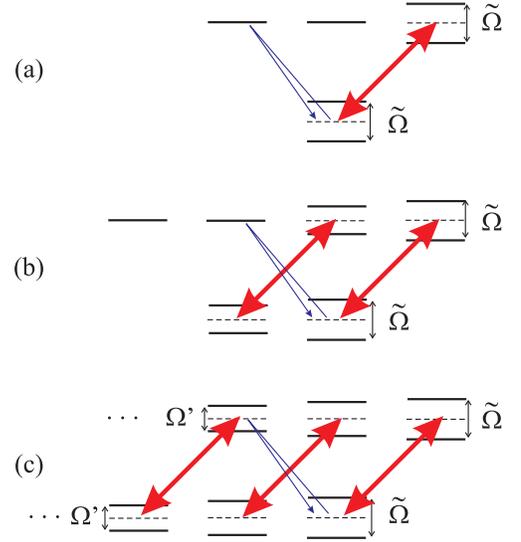}
\caption{(color online) Dressed state structure for atoms with different ground state angular momenta
(a) $J_g=0,$ (b) $ J_g = 1/2$, (c) $ J_g \geq 1$. Modified Rabi frequency $\tilde{\Omega}=\sqrt{\Omega^2+\delta^2}$, and $\Omega^\prime$ is given by Eq. (\ref{eq:omega_prime}). While in cases (a) and (b) the structure of the levels
relevant for double scattering is the same, in case (c) the excited state of the CBS transition is
dressed by the laser field.}
\label{fig:dressed_state}
\end{figure}
\begin{figure}[t]
\includegraphics[width=8cm]{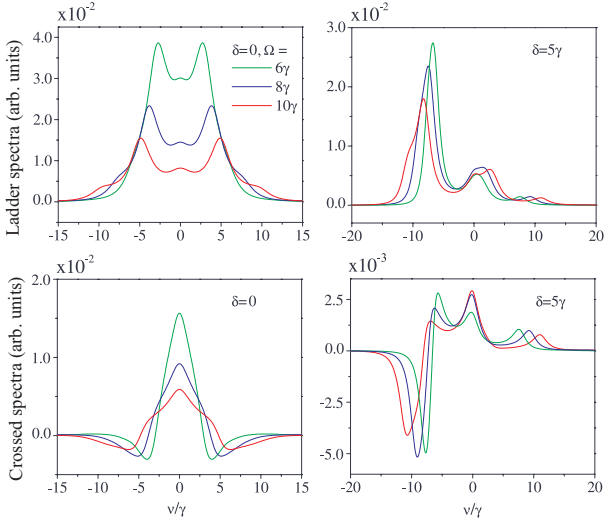}
\caption{(color online) Examples of double scattering ladder (top)
and crossed (bottom) spectra, for three different values of the Rabi
frequency $\Omega$ (see legend) at resonance ($\delta=0$; left) and
at a detuning $\delta = 5 \gamma$ (right), for $J_g=0$ and
$J_g=1/2$. The spectra for both values of $J_g$ coalesce upon
rescaling the $J_g= 1/2$ spectra by a factor $(4 J_g +1)^{-2}=1/9$.}
\label{fig:spectra_Sr}
\end{figure}
Figure \ref{fig:dressed_state} schematically depicts the dressed
levels of the optically pumped atoms with (a) $J_g=0$, (b)
$J_g=1/2$, and (c) $J_g\geq 1$. In the former two cases, the
structure of the dipole transition, relevant for the CBS signal in
the h $\parallel$ h channel, is the same. Unsurprisingly, the ladder
and crossed spectra for $J_g=0$ and $J_g=1/2$ plotted in
Fig.~\ref{fig:spectra_Sr} are identical, up to the numerical
prefactor $1/9=(4 J_g+1)^{-2}$ from Eq.~(\ref{eq:analytic_elastic}).
Note that, unlike Fig.~\ref{fig:weakly}, we present the ladder and
crossed spectra in the saturation regime in separate plots -- to
facilitate the interpretation of the spectral features which become
more complicated in this high intensity limit. Since the CBS spectra
for Sr atoms have been discussed in detail in \cite{shatokhin07}, we
right away move on to the case $J_g\geq 1$.

Results for the two examples $J_g=1$ and $J_g=3$ are presented
in Fig.~\ref{fig:spectra_Rb}. As in the weakly inelastic regime,
the spectra for $J_g\geq 1$ are different from those for $J_g< 1$:
Especially the number and the positions of the peaks differs.
The main reason for this distinction in the saturation regime is a different
dressed-state structure for $J_g\geq 1$ as compared to $J_g< 1$, see Fig.~\ref{fig:dressed_state}.

In the limit of well-separated spectral lines, $\tilde{\Omega}\gg \gamma$, the splitting between the
dressed levels corresponding to the transition $|J_g\,J_g\ra\leftrightarrow |J_e\,J_e\ra$ is equal to
the modified Rabi frequency $\tilde{\Omega}=\sqrt{\Omega^2+\delta^2}$, whereas the
splitting $\Omega^\prime$ between the dressed levels for the transition
$|J_g\,J_g\!-\!2\ra\leftrightarrow |J_e\,J_e\!-\!2\ra$ is given by the product of the modified
Rabi frequency and the corresponding Clebsch-Gordan coefficient,
\begin{align}
\Omega^\prime&=\tilde{\Omega}\la
J_g\, J_g\!\!-\!\!2,1 1|J_g\!\!+\!\!1\, J_g\!\!-\!\!1\ra\n\\
&=\tilde{\Omega}\sqrt{\frac{J_g(2 J_g-1)}{2J_g^2+3J_g+1}}.
\label{eq:omega_prime}
\end{align}
Due to these unequal splittings, there should appear four resonance frequencies in the CBS ladder spectra,
which represent the various double scattering processes, at
\be
\nu=\frac{1}{2}\lt(\pm \tilde{\Omega}\pm \Omega^\prime\rt).
\label{eq:dressed_resonances}
\e
Formula (\ref{eq:dressed_resonances}) describes accurately the positions of the resonances not only in
the limit of well-separated spectral lines, but also for moderate values of the Rabi frequency.
For instance, let us take $\delta=0$ and $\Omega=10\gamma$. In this case, the positions of
the maxima of the ladder spectrum as obtained in Figs.~\ref{fig:spectra_Rb}(a) and (e) (red lines) from the solution
of Eqs.~(\ref{eq:solution_gen_OBE}) and (\ref{eq:inel_blocks}) (with a binning size $0.1 \gamma$ of the frequency axis)
 are (in units of $\gamma$):
\begin{align*}
\nu&=\pm 2.8; \pm 7.1 \; (J_g=1),\\
\nu&=\pm 1.3; \pm 8.6\; (J_g=3).
\end{align*}
In good agreement with these values, Eqs.~(\ref{eq:omega_prime}) and (\ref{eq:dressed_resonances}) yield
resonances centered at the frequencies
\begin{align*}
\nu&=\pm 2.96; \pm 7.04 \; (J_g=1),\\
\nu&=\pm 1.34; \pm 8.66 \; (J_g=3).
\end{align*}
\begin{widetext}
\begin{center}
\begin{figure}
\includegraphics[width=17cm]{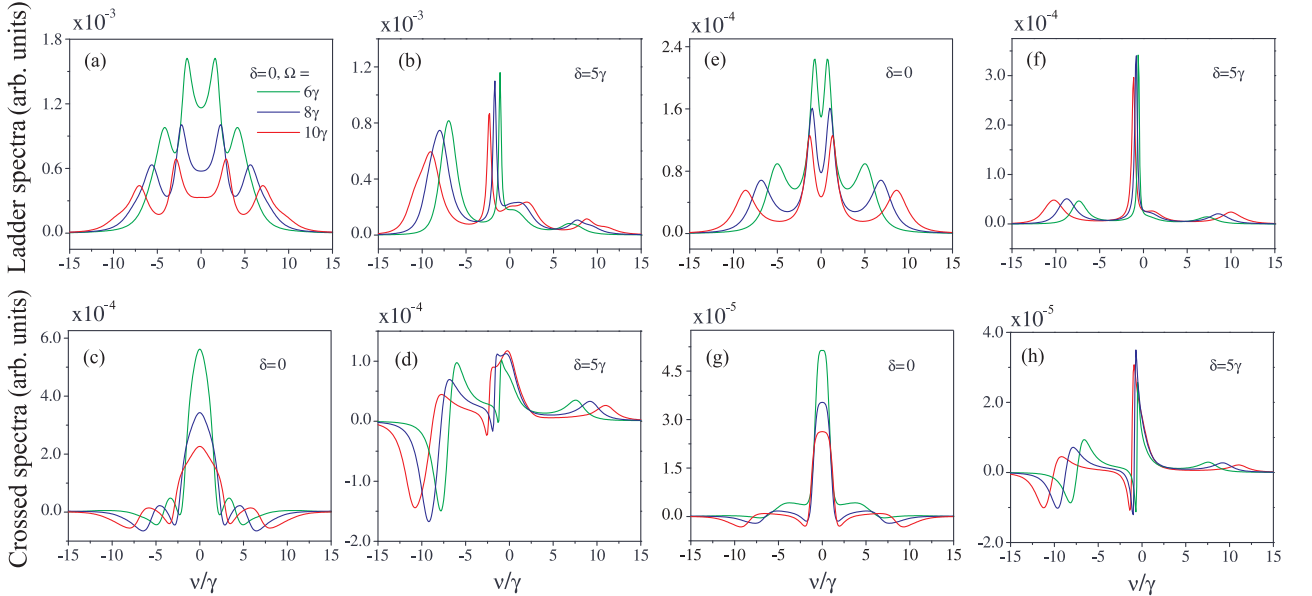}
\caption{(color online) Examples of the double scattering ladder (top) and crossed (bottom) spectra, for
three different values of the Rabi frequency $\Omega$ (see legend) for (a-d): $J_g=1$;
and (e-h): $J_g=3$. Plots (a), (c), (e), and (g) are obtained at exact resonance,  $\delta=0$, while (b), (d), (f),
and (h) show the result for finite detuning, $\delta=5\gamma$. }
\label{fig:spectra_Rb}
\end{figure}
\end{center}
\end{widetext}

As regards the crossed spectra, they originate from interferences
between different inelastic scattering processes that are manifest
in the ladder spectra as separate resonances \cite{shatokhin07}.
These interferences lead to a peculiar line shape of the crossed
spectra for $J_g=1$ and $J_g=3$ (see Fig.~\ref{fig:spectra_Rb})
which contains regions of, both, constructive and destructive
interference, depending on the phase shifts associated with the
corresponding frequency shifts upon inelastic scattering processes.
Note that, in all cases, the maximum of the crossed spectrum occurs
close to $\nu=0$. Therefore, the interference is always constructive
close to the laser frequency.
\begin{figure}
\includegraphics[width=8cm]{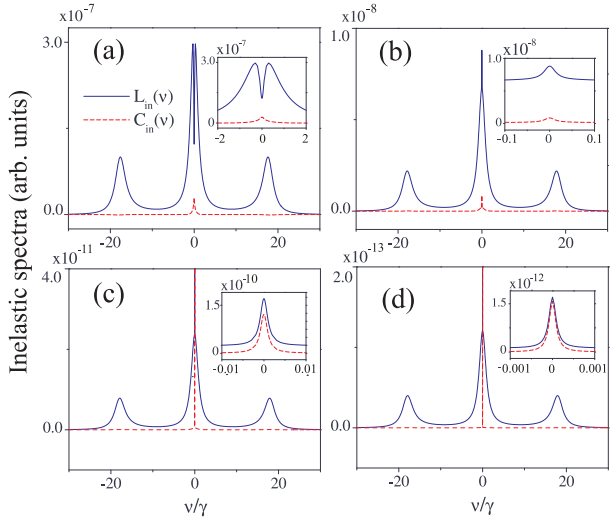}
\caption{(color online) Inelastic ladder (solid) and crossed
(dashed) double scattering CBS spectra at exact resonance ($\delta=0$) in the deep saturation regime ($\Omega=18\gamma$), for different dipole transitions: (a) $J_g=30$, (b) $J_g=110$, (c) $J_g=10^3$, and (d) $J_g=10^4$. Insets magnify narrow resonances.} \label{fig:spectra_Asymp_J}
\end{figure}

It is interesting to study the double scattering spectra for larger values of $J_g$. In particular, this will allow us, in the next section, to answer the question of whether the CBS interference effect survives in the limit $J_g\to\infty$. Previously, it was established that a residual enhancement factor exists in the deep saturation regime for atoms with $J_g=0$ \cite{shatokhin05}, and in the elastic scattering regime for semiclassical atoms ($J_g\to\infty$) \cite{cord_thesis}.

Using the aforementioned fact that optically pumped atoms with
arbitrary $J_g$ can be modeled as effective few-level systems,
it is possible to calculate the double scattering CBS spectra for
arbitrary $J_g$ by simply readjusting the values of the
Clebsch-Gordan coefficients. As we checked, these spectra look
qualitatively the same as the spectra shown in
Fig.~\ref{fig:spectra_Rb}, as long as $J_g\lesssim 40$. For
larger values of $J_g$, the splitting $\Omega^\prime$ between the dressed levels converges to the modified Rabi frequency $\tilde{\Omega}$ (see Eq.~(\ref{eq:omega_prime})), and the
two maxima of the ladder spectrum at $\nu=\pm
(\tilde{\Omega}-\Omega^\prime)/2$ merge. As a result, the ladder spectrum acquires a line shape consisting of three broad (linewidth $\sim \gamma$) peaks located at $\nu=-\tilde{\Omega}, 0, +\tilde{\Omega}$, and a subnatural linewidth peak at $\nu=0$,  see Fig.~\ref{fig:spectra_Asymp_J}. The three broad peaks represent nothing but the Mollow triplet \cite{mollow69}. It results from the spontaneous emission of the atom, excited by the probe field on the level $|J_e\,\, J_e\!-\!2\ra$, down the dressed states of the CBS transition (which tends, for $J_g\gg 1$, to the dressed-state structure of the laser-driven two-level atom, see Fig.~\ref{fig:dressed_state}(c)). As for the narrow resonance, we believe that it originates from destructive interference between two stimulated emission processes from the dressed states, when $\nu\approx 0$ (leading to an extremely long lifetime of these states $\sim (J_g /\gamma)$). We deduce the linear scaling with $J_g$ from the observation of the behavior of the widths of the subnatural peaks   which decrease as $\sim J_g^{-1}$ (see Fig.~\ref{fig:spectra_Asymp_J}). Similar ultranarrow spectral features due to destructive interference between the dressed state transitions were predicted in resonance fluorescence of a four-level atom excited by a bichromatic coherent field \cite{zhu99}.     

Concerning the crossed spectra, in the limit $J_g\gg 1$ it consists of a single positive narrow peak centered at $\nu=0$ (see Fig.~\ref{fig:spectra_Asymp_J}). Its width coincides with the width of the narrow ladder resonance and, hence, it also decreases as $\sim J_g^{-1}$ with increasing $J_g$.

To see better how the
above described spectral signatures affect the net interference effect of all the elastic and inelastic scattering processes, we conclude our
study with an investigation of the total CBS enhancement factor, in
the next section.

\subsection{Enhancement factor}
\label{sec:enhancement} The enhancement factor $\alpha$ is a
quantitative measure of phase coherence between the interfering
waves which contribute to the CBS signal. In the h $\parallel$ h
channel, it is defined as \cite{jonckheere00} \be
\alpha(\theta)=1+\frac{C_{\rm tot}(\theta)}{L_{\rm tot}},
\label{def:enhancement} \e where $\theta$ is the observation angle
with respect to the backwards direction, and $C_{\rm tot}(\theta)$
and  $L_{\rm tot}$ are the total crossed and ladder intensities of
double scattering, respectively. Hereafter, we consider the exact
backward direction, $\theta=0$.

In the inelastic scattering regime, the total intensities are given by the sums of the elastic and
inelastic intensities
\begin{align}
C_{\rm tot}(0)&=C_{\rm el}+C_{\rm in},\\
L_{\rm tot}&=L_{\rm el}+L_{\rm in},
\end{align}
where the elastic components, $C_{\rm el}=L_{\rm el}$, are defined in Eq.~(\ref{eq:analytic_elastic}),
and the inelastic intensities
\be
C_{\rm in}=\int_{\infty}^{\infty}d\nu C_{\rm in}(\nu), \quad L_{\rm in}=\int_{\infty}^{\infty}d\nu L_{\rm in}(\nu),
\label{eq:cin_lin}
\e
are given by integrations over their frequency distributions.
\begin{figure}
\includegraphics[width=7.5cm]{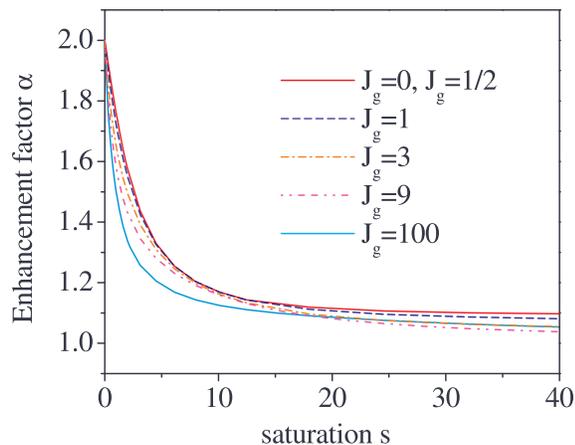}
\caption{(color online) Coherent backscattering enhancement factor
$\alpha$ vs. saturation $s$, at resonant driving, for different
values of $J_g$. The enhancement factor initially decreases faster
with increasing saturation for higher values of $J_g$.}
\label{fig:enhancement}
\end{figure}

Applying the formulas (\ref{def:enhancement})-(\ref{eq:cin_lin}) to the calculated spectra,
we study the behavior of the enhancement factor versus the saturation parameter, for different
values of the ground state angular momenta. Our results for the case of exact resonance are presented
in Fig.~\ref{fig:enhancement}.

 In the elastic scattering regime, that is for $s\to 0$, the enhancement factor features perfect
 phase coherence -- $\alpha\to 2$ -- for arbitrary $J_g$. This is in full agreement with our result for
 the elastic ladder and crossed intensities, see Eq.~(\ref{eq:analytic_elastic}).
 Furthermore, the results for $J_g=0$ and $J_g=1/2$ coincide for all $s$.
 As already discussed in Sec.~\ref{sec:inelastic}, the ground state degeneracy does not affect the
 phase coherence in these cases; the decrease of $\alpha$ is due to inelastic scattering processes alone.

Starting from $J_g=1$, the enhancement factor exhibits an initially
steeper decay of $\alpha$ with $s$ as $J_g$ increases. We attribute
this behavior to the fact that the coupling of the excited state
$|J_e\,J_e\!-\!2\ra$ to the ground state  $|J_g\,J_g\!-\!2\ra$
increases with $J_g$, due to the growth of the associated
Clebsch-Gordan coefficients.
Although, at intermediate and large values of $s$, larger values of
$J_g$ do not necessarily lead to a faster decrease of $\alpha$ with
$s$, the result for $J_g=0$ and $J_g=1/2$ yields an upper bound. In
other words, when the internal degeneracy comes into play, it always
leads to a faster decay of the phase coherence as compared to the
non-degenerate case.
\begin{figure}
\includegraphics[width=8cm]{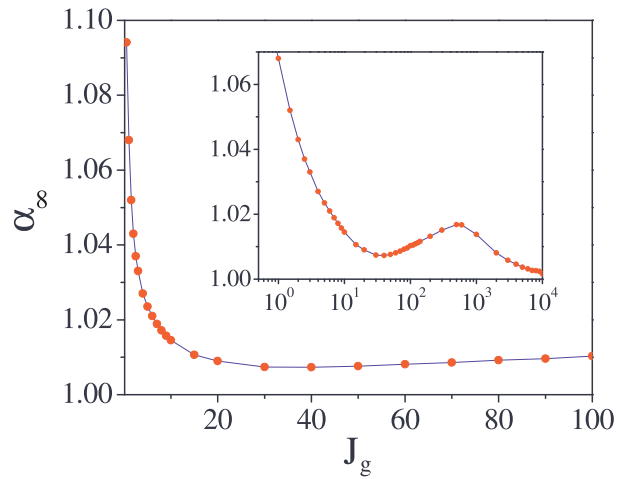}
\caption{(color online) Residual enhancement factor $\alpha_\infty$
at $s=162\gg 1$, versus $J_g=0$, 1/2, 1, 3/2, 2, 5/2, 3, 4, 5, 6, 7,
8, 9, 10, 15, etc, with $\alpha_\infty(0)=\alpha_\infty(1/2) \approx
1.095$. Inset: semi-log graph of $\alpha_\infty(J_g)$, with $\alpha_\infty(10^4)\approx 1.0016$. The continuous line
guides the eye.} \label{fig:enhancement1}
\end{figure}

Finally, let us discuss the asymptotic behavior of the enhancement
factor $\alpha_\infty$ in the deep saturation regime, $s\gg 1$. For
double scattering from Sr atoms, we found earlier that the inelastic
ladder and crossed intensities asymptotically decrease as $\sim
s^{-1}$, leading to a residual enhancement $\alpha_\infty\approx
1.095$ in the case of resonant driving \cite{shatokhin05}. The
dependence of $\alpha_\infty$ on $J_g$ is presented in
Fig.~\ref{fig:enhancement1}. For $J_g>1/2$, $\alpha_\infty$ drops
with increasing total angular momentum, until it reaches a minimum
of $\alpha_\infty(40)\approx 1.0073$. Further increase of $J_g$
leads to a very slow but monotonous increase of the residual
enhancement until $J_g\approx 500$, where a local maximum of $\alpha_\infty(500)\approx 1.017$ is reached (see inset in Fig.~\ref{fig:enhancement1}). This  behavior is unsurprising when taking into account
that the increase in $J_g$ is {\it not} accompanied by an increase
in the {\it effective} internal ground state degeneracy, which
remains equal to 2 for any $J_g\geq 1$. The slow growth  of
$\alpha_\infty$ for $J_g>40$ can be attributed to the fact that the total weight of the ladder spectrum decreases after merging two central peaks into one single peak (see Fig.~\ref{fig:spectra_Asymp_J}(a) and (b)). 

Further increase of $J_g$ leads to a monotonous decrease of $\alpha_\infty$. As follows from the discussion in Sec.~\ref{sec:saturated} and Fig.~\ref{fig:spectra_Asymp_J}, for very large values of $J_g$, an increase of $J_g$ is accompanied by narrowing of the subnatural linewidth resonances of the ladder and crossed spectra, without affecting the broad spectral features of the ladder spectra. This is not compensated by an increase of the relative peaks' heights, which remain fixed for a given value of $s$. Therefore, we predict that the enhancement factor should asymptotically tend to unity:
\be
\alpha_\infty= 1+O(J_g^{-1}), \quad J_g\to \infty. 
\e

\section{Summary and Conclusion}
\label{sec:conclusion} In this work, we generalized the pump-probe
approach to CBS of light by cold two-level atoms
\cite{geiger10,wellens10,shatokhin12a} to atoms with degenerate
energy levels. For this, we derived equations of motion for a
generalized Bloch vector, describing the dynamics of a single atom
under a classical bichromatic driving field. Because these equations
are formally equivalent to the equations appearing in the pump-probe
approach for two-level atoms, we could translate our equations to
the same diagrammatic language. By doing so, we obtained similar
single-atom building blocks as in \cite{shatokhin12a}, where, in the
generalized diagrams, each incoming and outgoing arrow is
additionally equipped with a polarization index. Like for two-level
atoms, the double scattering contributions to CBS can be derived by
combining these single-atom building-blocks self-consistently.

We applied the generalized pump-probe approach to study double
scattering from optically pumped atoms in the helicity preserving
polarization channel. To this end,  we considered several examples
of the dipole transition $J_g\rightarrow J_e=J_g+1$. Comparing our
results for the $J_g=0$ transition with the master equation results
\cite{shatokhin05,shatokhin07}, for different parameter values, we
could establish perfect agreement between both approaches.

For $J_g\geq 1$, the internal degeneracy manifests itself in the inelastic
scattering signal, leading to a faster decay of the CBS enhancement
factor with increasing saturation of the atomic transition as
compared to the non-degenerate case. Finally, we predict that, in the deep saturation regime, the CBS interference signal should asymptotically vanish with increasing $J_g$, as $J_g^{-1}$.

\acknowledgements
V.S. is grateful to I.M. Sokolov for engaging email correspondence and helpful remarks.
This work was financially supported by DFG, through grant BU-1337/9-1.

\appendix

\section{Projecting vectors}
\label{app:projections} For a dipole transition with the ground and
excited state angular momenta $J_g$ and $J_e$, respectively, the
complete orthogonal basis set contains $N=(2(J_e+J_g)+2)^2-1$
operators. We denote these operators by $\mu_1,\ldots, \mu_N$.
Consequently, the generalized optical Bloch vector can be written as
\be {\bf Q}=(\mu_1,\ldots,\mu_N)^T, \e where $T$ denotes
transposition. Among these operators, it is convenient to choose the
first $N_0=2(J_e+J_g)+2$ operators as the identity operator and
$N_0-1$ diagonal traceless operators. The remaining $N-N_0$
operators are chosen as non-diagonal operators describing
transitions between pairs of different sublevels. Then, the set of
operators $\mu^{\prime}_i$, orthogonal to the set of operators
$\mu_i$, can be chosen in the following way: The first $N_0$
operators of the orthogonal set read $\mu^{\prime}_i=\mu_i/{\rm
Tr}[\mu_i^2]$ ($i=1,\ldots,N_0$), and the remaining operators
$\mu_i^{\prime}=\mu_i^T$.

It is easy to see that, in this case, the orthogonality condition,
${\rm Tr}[\mu_n\mu_m^{\prime}]=\delta_{nm}$, is fulfilled for all
$1\leq n,m\leq N$. Consequently, any operator ${\cal O}=\sum_i
c_i\mu_i$ can be defined as a scalar product \be {\cal O}={\bf
C}\cdot{\bf Q}, \quad \la{\cal O}\ra={\bf C}\cdot\la{\bf Q}\ra, \e
where ${\bf C}=(c_1,\ldots,c_N)$ is a projecting vector, with
$c_i={\rm Tr}[{\cal O}\mu_i^{\prime}]$. Likewise, we denote the
vectors projecting onto the operator $D_q^\dagger$ and $D_q$ to be
${\bf U}_q$ and ${\bf V}_q$, respectively: \be \la
D_q^\dagger\ra={\bf U}_q\cdot\la {\bf Q}\ra, \quad\la D_q\ra={\bf
V}_q\cdot\la {\bf Q}\ra. \e

\section{Initial conditions}
\label{app:initial} We now explain how to define the initial
conditions in Eq. (\ref{eq:solf}). From the definitions of the
correlation functions (\ref{eq:inel_blocks}), we have \beml
\begin{align}
{\bf f}_{q}(0)&=\la {\bf Q}D_{q}\ra-\la{\bf Q}\ra\la D_{q}\ra,\label{f_q}\\
{\bf h}_{q^{\prime}}(0)&=\la D^\dagger_{q^{\prime}}{\bf Q}\ra-\la
D^\dagger_{q^{\prime}}\ra\la{\bf Q}\ra,\label{h_q}
\end{align}
\label{eq:initial_fluct}
\eml
where the average should be taken with respect to the steady state of a single laser driven atom.
We note that the perturbative expansion of the factorized part of the correlation function
in Eq.~(\ref{eq:initial_fluct}) can be obtained directly from Eq.~(\ref{eq:solution_gen_OBE_d}).
As regards the non-factorized parts on the right hand sides of (\ref{eq:initial_fluct}), they can be expressed using Eq.~(\ref{eq:solution_gen_OBE_d}) as follows
\beml
\begin{align}
\la {\bf Q}D_{q}\ra&={\bf A}_1\la{\bf Q}\ra+{\bf L}_1,\label{a_1}\\
\la D^{\dagger}_{q^{\prime}}{\bf Q}\ra&={\bf A}_2\la{\bf Q}\ra+{\bf
L}_2,\label{a_2}
\end{align}
\eml
where
\beml
\begin{align}
({\bf A}_1)_{ij}&={\rm Tr}[\mu_iD_{q}\mu^{\prime}_j], \quad ({\bf L}_1)_i={\rm Tr}[D_{q}\mu^{\prime}_i]/N_0,\\
({\bf A}_2)_{ij}&={\rm Tr}[D^\dagger_{q^\prime}\mu_i\mu^{\prime}_j],
\quad ({\bf L}_2)_i={\rm
Tr}[D^{\dagger}_{q^{\prime}}\mu^{\prime}_i]/N_0.
\end{align}
\eml

Performing the perturbative expansion of both sides of
Eq.~(\ref{f_q}) to second order in the probe field, we obtain:
\begin{widetext}
\beml
\begin{align}
{\bf f}^{(0)}_{q}(0)&={\bf A}_1\la{\bf Q}\ra^{(0)}+{\bf L}_1-
\la{\bf Q}\ra^{(0)}\la D_{q}\ra^{(0)},\\
{\bf f}^{(+)}_{q}(\omega^{[r^\prime]};0)&={\bf A}_1\la{\bf
Q}(\omega^{[r^\prime]})\ra^{(+)}-
\la{\bf Q}(\omega^{[r^\prime]})\ra^{(+)}\la D_{q}\ra^{(0)}-\la{\bf Q}\ra^{(0)}\la D_{q}(\omega^{[r^\prime]})\ra^{(+)},\\
{\bf f}^{(-)}_{q}(\omega^{[r]};0)&={\bf A}_1\la{\bf Q}(\omega^{[r]})\ra^{(-)}-\la{\bf Q}(\omega^{[r]})\ra^{(-)}\la D_{q}\ra^{(0)}-\la{\bf Q}\ra^{(0)}\la D_{q}(\omega^{[r]})\ra^{(-)},\\
{\bf f}_{q}^{(+-)}(\omega^{[r^\prime]},\omega^{[r]};0)&={\bf A}_1\la{\bf Q}(\omega^{[r^\prime]},\omega^{[r]})\ra^{(+-)}-\la{\bf Q}(\omega^{[r^\prime]},\omega^{[r]})\ra^{(+-)}\la D_{q}\ra^{(0)}\n\\&
-\la{\bf Q}\ra^{(0)}\la D_{q}(\omega^{[r^\prime]},\omega^{[r]})\ra^{(+-)}-\la{\bf Q}(\omega^{[r^\prime]})\ra^{(+)}\la D_{q}(\omega^{[r]})\ra^{(-)}\n\\
&-\la{\bf Q}(\omega^{[r]})\ra^{(-)}\la
D_{q}(\omega^{[r^\prime]})\ra^{(+)}.
\end{align}
\eml
\end{widetext}
Expanding, in the same way, Eq.~(\ref{h_q}) leads to the initial
conditions for the vector ${\bf h}_{q^\prime}(0)$:
\begin{widetext}
\beml
\begin{align}
{\bf h}^{(0)}_{q^\prime}(0)&={\bf A}_2\la{\bf Q}\ra^{(0)}+{\bf L}_2-\la{\bf Q}\ra^{(0)}\la D^{\dagger}_{q^\prime}\ra^{(0)},\\
{\bf h}^{(+)}_{q^\prime}(\omega^{[r^\prime]};0)&={\bf A}_2\la{\bf Q}(\omega^{[r^\prime]})\ra^{(+)}
-\la{\bf Q}(\omega^{[r^\prime]})\ra^{(+)}\la D^{\dagger}_{q^\prime}\ra^{(0)}
-\la{\bf Q}\ra^{(0)}\la D^{\dagger}_{q^\prime}(\omega^{[r^\prime]})\ra^{(+)},\\
{\bf h}^{(-)}_{q^\prime}(\omega^{[r]};0)&={\bf A}_2\la{\bf Q}(\omega^{[r]})\ra^{(-)}
-\la{\bf Q}(\omega^{[r]})\ra^{(-)}\la D^{\dagger}_{q^\prime}\ra^{(0)}
-\la{\bf Q}\ra^{(0)}\la D^{\dagger}_{q^\prime}(\omega^{[r]})\ra^{(-)},\\
{\bf h}_{q^\prime}^{(+-)}(\omega^{[r^\prime]},\omega^{[r]};0)&={\bf A}_2\la{\bf Q}(\omega^{[r^\prime]},\omega^{[r]})\ra^{(+-)}
-\la{\bf Q}(\omega^{[r^\prime]},\omega^{[r]})\ra^{(+-)}\la D^{\dagger}_{q^\prime}\ra^{(0)}\n\\
&-\la{\bf Q}\ra^{(0)}\la D^{\dagger}_{q^\prime}(\omega^{[r^\prime]},\omega^{[r]})\ra^{(+-)}
-\la{\bf Q}(\omega^{[r^\prime]})\ra^{(+)}\la D^{\dagger}_{q^\prime}(\omega^{[r]})\ra^{(-)}\n\\
&-\la{\bf Q}(\omega^{[r]})\ra^{(-)}\la
D^{\dagger}_{q^\prime}(\omega^{[r^\prime]})\ra^{(+)}.
\end{align}
\eml
\end{widetext}






\end{document}